%% file: main.tex
\documentclass[11pt]{article}

\usepackage{amsmath}
\usepackage{amssymb}
\usepackage{amsthm}
\usepackage[margin=1.25in]{geometry}
\usepackage{subfig}
\usepackage{graphicx}
\usepackage{xcolor}
\usepackage[bookmarksopen=true,bookmarksnumbered=true,colorlinks=true,linkcolor=black,anchorcolor=black,citecolor=black,urlcolor=black]{hyperref}
\usepackage{./common}

\numberwithin{equation}{section}

\newtheorem{theorem}{Theorem}[section]
\newtheorem{lemma}[theorem]{Lemma}
\newtheorem{corollary}[theorem]{Corollary}
\newtheorem{observation}[theorem]{Observation}

\title{\bf{Approximate Self-Assembly of the\\Sierpinski Triangle}\thanks{This research was supported in part by NSF grants 0652569 and 0728806.}}
\author{
	Jack H. Lutz\footnote{Department of Computer Science, Iowa State University, \url{lutz@cs.iastate.edu}} 
	\and 
	Brad Shutters\footnote{Department of Computer Science, Iowa State University, \url{shutters@cs.iastate.edu}} 
}
\date{}

\begin{document}
\maketitle
\input{abstract.tex}
\newpage
\input{intro.tex}

\input{prelim.tex}

\input{limits.tex}

\input{cd.tex}
\input{laced.tex}

\input{open.tex}

\section*{Acknowledgments}
We thank Jim Lathrop, Xiaoyang Gu, Scott Summers, Dave Doty, Matt Patitz, and Brian Patterson for useful discussions.
\bibliography{../../bibliography}
\end{document}

%% file: abstract.tex

\begin{abstract}
The Tile Assembly Model is a Turing universal model that Winfree introduced in order to study the nanoscale self-assembly of complex (typically aperiodic) DNA crystals.
Winfree exhibited a self-assembly that tiles the first quadrant of the Cartesian plane with specially labeled tiles appearing at exactly the positions of points in the Sierpinski triangle.
More recently, Lathrop, Lutz, and Summers proved that the Sierpinski triangle cannot self-assemble in the \dquote{strict} sense in which tiles are not allowed to appear at positions outside the target structure.
Here we investigate the strict self-assembly of sets that approximate the Sierpinski triangle.
We show that every set that does strictly self-assemble disagrees with the Sierpinski triangle on a set with fractal dimension at least that of the Sierpinski triangle ($\approx 1.585$), and
that no subset of the Sierpinski triangle with fractal dimension greater than 1 strictly self-assembles.
We show that our bounds are tight, even when restricted to supersets of the Sierpinski triangle, by presenting
a strict self-assembly that adds communication fibers to the fractal structure without disturbing it.
To verify this strict self-assembly we develop a generalization of the local determinism method of Soloveichik and Winfree.
\end{abstract} 

%% file: intro.tex

\section{Introduction}

Self-assembly is a process in which simple objects autonomously combine to form complex structures as a consequence of specific, local interactions among the objects themselves.
It occurs spontaneously in nature as well as in engineered systems and is a fundamental principle of structural organization at all scales. Since the pioneering work of Seemen \cite{najl}, the self-assembly of DNA molecules has developed into a field with rich interactions between the theory of computing (the information processing properties of DNA) and geometry (the structural properties of DNA), and with many applications to nanotechnology \cite{seeman2003a}.

Winfree \cite{winfreephd} introduced the Tile Assembly Model (TAM) as a mathematical model of DNA self-assembly in order to study the nanoscale self-assembly of complex (typically aperiodic) DNA crystals. 
It is a constructive version of Wang tiling \cite{wang1961a,wang1962a} that models the self-assembly of unit square \emph{tiles}
that can be translated, but not rotated. 
A tile has a \emph{glue} on each side that
is made up of a \emph{color} and an integer \emph{strength} (usually 0, 1, or 2).
Intuitively, a tile models a DNA double crossover molecule and the glues correspond to the \dquote{sticky ends} on the four arms of the molecule.
Two tiles with the same glue on each side are of the same \emph{tile type}.
Two tiles placed next to each other \emph{interact} if the glues on their abutting sides match in both color and strength.
A \emph{tile assembly system} (TAS) is a finite set of tile types, a single tile for the \emph{seed}, and a specified integer \emph{temperature} (usually 2).
The process starts with the seed tile placed at the origin and growth occurs by single tiles attaching one at a time.
A tile can attach at a site where the summed  strength of the glues on sides that interact with the existing structure is at least the temperature.
The assembly is \emph{terminal} when no more tiles can attach.
A TAS is \emph{directed} if it always results in a unique terminal assembly.
Winfree proved the TAM is Turing universal \cite{winfreephd}.
The TAM is described formally in Section \ref{sec:tam}.

This paper is concerned with the self-assembly of fractals. 
Structures that self-assemble in naturally occurring biological systems are often fractals of low dimension, which have advantages for materials transport, heat exchange, information processing, and robustness \cite{ssadst}.
Fractals are normally bounded and have
 the same detail at arbitrarily small scales.
But, the TAM models the bottom-up self-assembly of tiles which are discrete objects. So, structures that self-assemble in the TAM are fundamentally discrete. Thus, we consider the self-assembly of discrete fractals which are unbounded and have the same detail at arbitrarily large scales.
There are two main notions of the self-assembly of a fractal. In \emph{weak self-assembly}, one typically causes a two-dimensional surface to self-assemble with the desired fractal structure appearing as a labeled subset of the surface. In contrast, \emph{strict self-assembly} requires only the fractal structure, and nothing else, to self-assemble. For many purposes, strict self-assembly is needed in order to achieve the above mentioned advantages of fractal structures.

The Sierpinski triangle is a canonical \dquote{toy} problem for self-assembly. Winfree \cite{winfreephd} showed that the Sierpinski triangle weakly self-assembles, and Rothemund, Papadakis, and Winfree \cite{asadst} achieved a molecular implementation of this self-assembly.
Lathrop, Lutz, and Summers \cite{ssadst} proved that the Sierpinski triangle cannot strictly self-assemble.
Patitz and Summers \cite{sadssf} exhibited a large class of fractals that cannot strictly self-assemble.
It is an open question whether any self-similar fractal strictly self-assembles.
Thus, techniques are needed to approximate self-similar fractals with strict-self-assembly.
The only previously known technique, introduced by Lathrop, Lutz, and Summers \cite{ssadst},
and generalized by Patitz and Summers \cite{sadssf},
enables strict self-assembly by adding communication fibers
that shift successive stages of the fractal causing the result to only visually resemble, but not contain,
the intended fractal structure.

In this paper we address a quantitative question: given that the Sierpinski triangle $\S$ cannot strictly self-assemble, how closely can strict self-assembly approximate $\S$?
That is, if $X$ is a set that \emph{does} strictly self-assemble, how small can the fractal dimension of the symmetric difference $X\Delta\,\S$ be?  Our first main theorem says that the fractal dimension of $X\Delta\,\S$  is at least the fractal dimension of $\S$. To gain further insight, we restrict our attention to subsets of $\S$ and show that here the limitation is even more severe.  Any subset of the Sierpinski triangle that strictly self-assembles must have fractal dimension 0 or 1. Roughly speaking, the axes that bound $\S$ form the largest subset of $\S$ that strictly self-assembles.
Hence,  $\S$ cannot even be approximated \dquote{closely} with strict self-assembly.

Our second main theorem shows that our first main theorem is tight, even when restricted to supersets of $\S$.
To prove this we demonstrate the existance of a set $X$
with the following three properties.
\begin{itemize}
\item[(1)] $\S \subseteq X$.
\item[(2)] The fractal dimension of $X\Delta\,\S$ is the fractal dimension of $\S$.
\item[(3)] $X$ strictly self-assembles in the Tile Assembly Model.
\end{itemize}
What we have achieved here is a means of \emph{fibering $\S$ in place}, i.e., adding the needed communication fibers (the set $X\setminus\S$) without disturbing the set $\S$.

The local determinism method of Soloveichik and Winfree \cite{csas} is a common technique for proving a TAS is directed.
However, the TAS in the proof of our second main theorem uses a blocking technique that prevents it from being locally deterministic.
We thus introduce \emph{conditional determinism}, a generalization of local determinism, to verify this TAS is directed.

The proof techniques used here, along with our blocking technique (and thus our generalization of local determinism), are likely to be useful in the design and analysis of other tile assembly systems that approximate self-similar fractals. Our fibering technique may be a useful example for other contexts where one seeks to enhance the \dquote{internal bandwidth} of a set in a distortion-free manner. We hope that our results lead to a more general understanding of the limitations of self-assembly and the approximate self-assembly of self-similar fractals.

%% file: prelim.tex

\section{Preliminaries}
\input{notation.tex}
\input{tam.tex}

\input{zetadim.tex}
\input{sierpinski.tex}

%% file: notation.tex

\subsection{Notation and Terminology}

We work in the discrete Euclidean plane $\Ints[2]$. We write $U_2$ for the set of all \emph{unit vectors} in $\Ints[2]$. We often refer to the elements of $U_2$ as the cardinal directions, and write $\un$ for $(0,1)$, $\us$ for $(0,-1)$, $\ue$ for $(1,0)$, and $\uw$ for $(-1,0)$.

Let $X$ and $Y$ be sets. We write $[X]^2$ for the set of all $2$-element  subsets of $X$.
For a partial function $f:X \dashrightarrow Y$, we write $f(x)\!\downarrow$ if $x \in \dom f$ and $f(x)\!\uparrow$ otherwise.
We write $X \Delta\, Y$ for the \emph{symmetric difference} of $X$ and $Y$.
For a \emph{Boolean} expression $\phi$, $\bool{\phi}=1$ if $\phi$ is true, and $\bool{\phi}=0$ otherwise.

All graphs here are undirected graphs of the form $G=(V,E)$, where $V\subseteq\Ints[2]$ is a set of \emph{vertices} and $E\subseteq [V]^2$ is a set of \emph{edges}.
A \emph{grid graph} is a graph where each $\set{\vec{m}, \vec{n}} \in E$ satisfies $\vec{m}-\vec{n}\in U_2$.
If $E$ contains every $\set{\vec{m}, \vec{n}}\in [V]^2$ such that $\vec{m}-\vec{n} \in U_2$, we say it is the \emph{full grid graph} on $V$, written $G^{\#}_V$.
A \emph{cut} of a graph is a partition of $V$ into two subsets.
A \emph{binding function} on a graph is a function $\beta:E\rightarrow\Nats$.
If $\beta$ is a binding function on $G$ and $C$ is a cut of $G$, then the \emph{binding strength of $\beta$ on $C$} is $$\beta_C = \sum\set{\beta(e) \mid e \in E \text{ and } e \cap C_0 \ne \emptyset \text{ and } e \cap C_1 \ne \emptyset},$$ and the \emph{binding strength of $\beta$ on $G$} is $$\min{\set{\beta_C \mid C \text{ is a cut of }G}}.$$
A \emph{binding graph} is an ordered triple $(V,E,\beta)$, where $\beta$ is a binding function on $(V,E)$.
For $\tau \in \Nats$, a binding graph $(V,E,\beta)$ is \emph{$\tau$-stable} when $\beta_{(V,E)} \ge \tau$.

We now review finite-tree depth \cite{ssadst}. Let $G\!=\!(V,E)$ be a graph and let $D \subseteq V$.
 For $r \in V$, the \emph{$D$-$r$-rooted subgraph} of $G$ is the graph $G_{D,r}\!=\!(V_{D,r},E_{D,r})$, where
\begin{align*}
	V_{D,r} = \set{v \in V \mid \text{$r$ is on every path from $v$ to (any vertex in) $D$}}
\end{align*}
and $E_{D,r} = E \cap [V_{D,r}]^2$.
 A \emph{$D$-subtree} of $G$ is a rooted tree $B$ with root $r \in V$ such that $B=G_{D,r}$.
The \emph{finite-tree depth} of $G$ \emph{relative to} $D$ is
\begin{equation*}
	\ftdepth{D}(G) = \sup \set{\depth(B) \mid \text{$B$ is a finite $D$-subtree of $G$}}.
\end{equation*}
Intuitively, given a set $D$ of vertices of $G$ (which is in practice the domain of the seed assembly), the $D$-subtree of $G$ is a rooted tree in $G$ that consists of all vertices of $G$ that lie at or on the far side of the root from $D$.

%% file: tam.tex
\vspace{-1pt}
\subsection{The Tile Assembly Model}\label{sec:tam}
\vspace{-1pt}
We now review the Tile Assembly Model \cite{winfreephd,rothemundphd,pscsas}. Our notation follows that of \cite{ssadst} but is tailored somewhat to our objectives.

\begin{figure}[!ht]
 	\centering
    	\captionsetup{margin=18pt}
 	\includegraphics{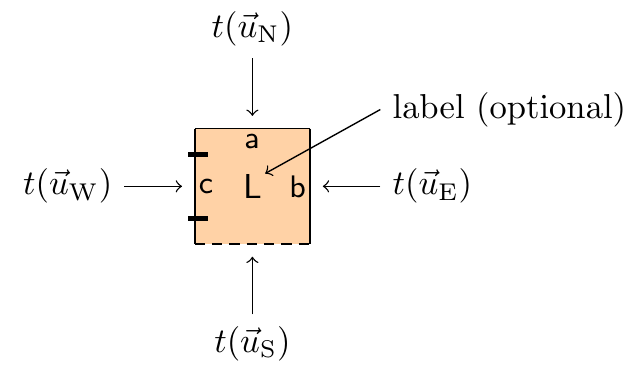}
	\caption{An example illustration of a tile. Glue strengths are represented by lines that are dotted for 0, solid for 1, and solid with notches for 2. Glue colors are drawn in the interior of the tile on the corresponding side. In this example $t(\un)=(a,1)$, $t(\ue)=(b,1)$, $t(\us)=(\lambda,0)$ where $\lambda$ represents the empty string, and $t(\uw)=(c,2)$. Also, we sometimes give a label to a tile type. This label does not play a role in the TAM, it is only to make referring to tiles of that type more convenient.}
	\label{fig:Tile}
\end{figure}

A \emph{tile} $t$ is a unit square that can be translated, but not rotated, so it has a well defined \dquote{side $\vu$\,} for each $\vu \in U_2$.
Each side $\vu$ of $t$ has a \emph{glue} $t(\vu)=(\gcol(\vu), \gstr(\vu))$ where $\gcol(\vu)\in\Sigma^*$, for some fixed alphabet $\Sigma$, is the glue \emph{color}, and $\gstr(\vu)\in\Nats$ is the glue \emph{strength}.
Two tiles with the same glue on each side are of the same \emph{tile type}. See Figure \ref{fig:Tile} for an example illustration of a tile.

Let $T$ be a set of tile types. A $T$-configuration is a partial function $\alpha : \Ints[2] \dashrightarrow T$.
For $\vm, \vn \in \dom\alpha$, the tiles at these locations \emph{interact} with \emph{strength} $$\gstr[\alpha](\vm,\vn) = \bool{\vn - \vm \in U_2} \cdot \gstr[\alpha(\vm)](\vn-\vm) \cdot \bool{\alpha(\vm)(\vu)=\alpha(\vn)(-\vu)}.$$
The \emph{binding graph} of $\alpha$ is $G_{\alpha}=(\dom\alpha,E,\beta)$, where $$E =\set{\{\vm,\vn\} \in [V]^2 \mid \gstr[\alpha](\vm,\vn)>0},$$ and for all $\set{\vm,\vn} \in E$, $\beta(\set{\vm,\vn}) = \gstr[\alpha](\vm,\vn).$
For $\tau\in\Nats$, $\alpha$ is \emph{$\tau$-stable} if $G_{\alpha}$ is $\tau$-stable.
We write $\assemblies$ for the set of all $\tau$-stable $T$-configurations.
Let $\alpha, \alpha' \in \assemblies$. If $\dom\alpha \subseteq \dom\alpha'$ and $\alpha(\vm)=\alpha'(\vm)$ for all $\vm \in \dom\alpha$, then $\alpha$ is a \emph{subconfiguration} of $\alpha'$ and we write $\alpha \sqsubseteq \alpha'$. If $|\dom\alpha'\setminus\dom\alpha|=1$, then $\alpha'$ is a \emph{single-tile-extension} of $\alpha$ and we write $\alpha' = \alpha + (\vm \mapsto t)$ where $\set{\vm} = \dom\alpha'\setminus\dom\alpha$ and $t=\alpha'(\vm)$.
For each $t \in T$, the \emph{$\tau$-$t$-frontier of $\alpha$} is
\begin{displaymath}
    \tfrontier{t}\alpha =
		\set{
    			\vm \in \Ints[2]\!\setminus\!\dom\alpha
    		\mid
			(	
				\Sigma_{\vu \in U_2}\,
					\gstr[\alpha + (\vm \mapsto t)](\vm,\vm+\vu)
			)	
			\ge
			\tau
		}, \text{ and}
\end{displaymath}
the \emph{$\tau$-frontier of $\alpha$} is  $$\frontier\alpha = \bigcup_{t \in T}\tfrontier{t}\alpha.$$ We say $\alpha$ is \emph{terminal}  when $\frontier\alpha=\emptyset$.

A \emph{tile assembly system} (TAS) is an ordered triple $\calT\!=\!(T, \sigma, \tau)$ where $T$ is a finite set of tile types,  the \emph{seed assembly} $\sigma\!\in\!\assemblies$ is such that $\dom\sigma\!=\!\vzero$, and $\tau\!\in\!\Nats$ is the \emph{temperature}.
An \emph{assembly sequence in $\calT$} is a sequence $\valpha\!=\!(\alpha_i \mid 0 \le i < k)$ where $\alpha_0\!\in\!\assemblies$, $k\!\in\!\Ints[+]\!\cup\!\set{\infinity}$ and for each
$0\!\le\!i\!<\!k$, $\alpha_{i+1}\!=\!\alpha_i + (\vm \mapsto t)$
for some $t\!\in\!T$ and $\vm\!\in\!\tfrontier{t}\alpha_i$.
The \emph{result of $\valpha$}, written $\res\valpha$, is the unique $\alpha\in\assemblies$ satisfying $\dom\alpha\!=\!\bigcup_{0 \le i < k}\!\dom\alpha_i$ and for each $0\!\le\!i\!<\!k$, $\alpha_i\!\sqsubseteq\!\alpha$.
We write $\alpha_0\!\extension\!\res\valpha$ if an assembly sequence from $\alpha_0$ to $\res\valpha$  exists. The set of \emph{producible assemblies} is $\producibles\!=\!\set{\alpha\!\in\!\assemblies \mid \sigma\!\extension\!\alpha}$ and the set of \emph{terminal assemblies} is $\terminals\!=\!\set{\alpha \in \producibles \mid \frontier \alpha\!=\!\emptyset}$.
$\calT$ is \emph{directed} if $|\terminals|\!=\!1$.
A set \emph{$X$ strictly self-assembles in $\calT$} if every $\alpha\!\in\!\terminals$ satisfies
$\dom\alpha\!=\!X$.  We say $X$ \emph{strictly self-assembles} if $X$ strictly self-assembles in some TAS.

Let $\calT=(T,\sigma,\tau)$ be a TAS, $\valpha = (\alpha_i \mid 0 \le i < k)$ be an assembly sequence in $\calT$, and $\alpha=\res\valpha$.
For each $\vm\in\Ints[2]$, the \emph{$\valpha$-index} of $\vm$ is $i_{\valpha}(\vm)=\min\set{i \in \Nats \mid \vm \in \dom\alpha_i}$.
If $\vm, \vn \in \dom\alpha$ and $i_{\valpha}(\vm) < i_{\valpha}(\vn)$, we say \emph{$\vm$ precedes $\vn$ in $\valpha$}, and write $\vm \prec_{\valpha} \vn$.
For $X\subseteq\dom\alpha$,
\emph{$\alpha$ restricted to $X$}, written
$\alpha \restrictedto X$, is the unique $T$-configuration satisfying $(\alpha\!\restrictedto\!X)\!\sqsubseteq\!\alpha$ and $\dom(\alpha\!\restrictedto\!X)\!=\!X$.

Winfree and Soloveichik \cite{csas} introduced local determinism as a convenient way to prove a TAS is directed.
Let $\calT=(T,\sigma,\tau)$ be a TAS, $\valpha = (\alpha_i \mid 0 \le i < k)$ be an assembly sequence in $\calT$, and $\alpha=\res\valpha$.
For each $\vm \in \dom\alpha$, define \cite{csas} the sets
\begin{align*}
\insides(\vm) &= \{\vu \in U_2 \mid \vm + \vu \prec_{\valpha} \vm \tand\ \gstr[\alpha_{i_{\valpha}(\vm)}](\vm,\vm+\vu) > 0\}, \text{ and}\\
\outsides(\vm) &= \{\vu \in U_2 \mid -\vu \in \insides(\vm+\vu)\}.
\end{align*}
Then, $\valpha$ is \emph{locally deterministic} \cite{csas} if the following three conditions hold.
\begin{itemize}
\item[(1)] For all $\vm \in \dom\alpha \setminus \dom\alpha_0$, $$\sum_{\vu\in\insides(\vm)} \gstr[\alpha_{i_{\valpha}(\vm)}](\vm,\vm+\vu) = \tau.$$
\item[(2)] For all $\vm \in \dom\alpha \setminus \dom\alpha_0$ and $t \in T \setminus \set{\alpha(\vm)}$,
	  $$
	  	\vm \not\in \tfrontier{t} (\alpha \upharpoonright (\dom\alpha \setminus (\set{\vm} \cup (\vm + \outsides(\vm))))).
	  $$
\item[(3)] $\frontier \alpha =\emptyset$.
\end{itemize}
Conceptually, (1) requires that each tile added in $\valpha$ \dquote{just barely} binds to the existing assembly; (2) holds when the tiles at $\vm$ and $\vm + \outsides(\vm)$ are removed from $\alpha$, no other tile type can attach to the assembly at location $\vm$; and (3) requires that $\alpha$ is terminal.
A TAS is \emph{locally deterministic} if it has a locally deterministic assembly sequence.
Soloveichik and Winfree \cite{csas} proved every locally deterministic TAS is directed.

%% file: zetadim.tex

\subsection{{Zeta-Dimension}}\label{sec:zetadim}

A fractal dimension is a measure of how completely a fractal fills space. The most commonly used fractal dimension for discrete fractals is zeta-dimension. 
Although the origins of zeta-dimension lie in eighteenth and nineteenth century number theory, namely Euler's \emph{zeta-function} \cite{euler1737a}, it has been rediscovered many times by researchers in a variety of fields. 
See \cite{zetadim} for a review of the origins of zeta-dimension, the development of its basic theory, and the connections between zeta-dimension and classical fractal dimensions.

In this paper we use the entropy characterization of zeta-dimension \cite{cahen1894a}.
For each $\vec{m} \in \Ints[2]$, let $||\vec{m}||$ be the Euclidean distance from the origin to $\vec{m}$, i.e., if $\vec{m}=(m_1,m_2)$ then $||\vec{m}||=\sqrt{m_1^2 + m_2^2}$. 
For $A \subseteq \Ints[2]$ and $I \subseteq [0,\infinity)$, let $A_I = \set{\vec{m} \in A \mid ||\vec{m}|| \in I}$. 
Then, the \emph{$\zeta$-dimension} (\emph{zeta-dimension}) of a set $A \subseteq \Ints[2]$ is
\begin{equation*}
    \ZetaDim(A) = \limsup_{n \rightarrow \infty} \frac{\log_2{|A_{[0,n]}|}}{\log_2{n}}.
\end{equation*}
By routine calculus it follows that
\begin{equation}\label{eqn:zetadim}
    \ZetaDim(A) = \limsup_{n \rightarrow \infty} \frac{\log_2{|A_{[0,2^n)}|}}{n}.
\end{equation}
 
Note that $\zeta$-dimension has the following functional properties of a fractal dimension \cite{zetadim}.

\begin{observation}\label{obs:ZetaDimProps}
Let $A,B \subseteq \Ints[2]$. Then,
\begin{itemize}
\item[(1)] $A \subseteq B \implies \ZetaDim(A) \le \ZetaDim(B)$ (monotonicity), and
\item[(2)] $\ZetaDim(A \cup B) = \max\set{\ZetaDim(A),\ZetaDim(B)}$ (stability).
\end{itemize}
\end{observation}

%% file: sierpinski.tex

\subsection{The Sierpinski Triangle}\label{sec:sierpinski}

\begin{figure}[h!t!]
	\centering
	\includegraphics{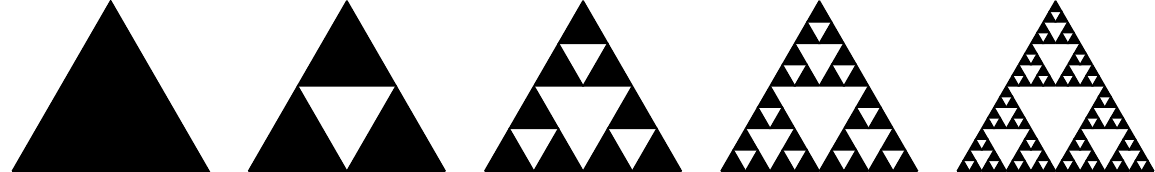}
	\caption{The first five stages of the continuous Sierpinski triangle.}
	\label{fig:continuous}
\end{figure}

The Sierpinski triangle, a.k.a. the Sierpinski gasket or the Sierpinski sieve, is a self-similar fractal named after the Polish mathematician Waclaw Sierpi\'{n}ski who first described it \cite{sierpinski}. It is formed by starting with a solid triangle and removing the middle fourth. This process is continued ad infinitum on all remaining triangles. See Figure \ref{fig:continuous} for an illustration of the this process.

This continuous version of the Sierpinski triangle is bounded and has the same detail at arbitrarily small scales.  But, because the TAM models the bottom-up self-assembly of tiles, which are discrete objects, structures that self-assemble in the TAM are fundamentally discrete. Therefore, we shall focus on the strict self-assembly of a discrete version of the Sierpinski triangle that is unbounded and has the same detail at arbitrarily large scales.

\begin{figure}[h!t!]
	\centering
	\includegraphics{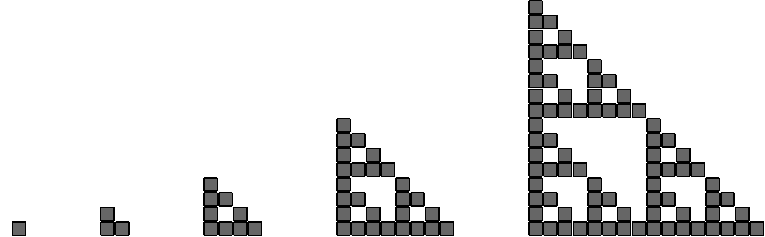}
	\caption{The first five stages of the discrete Sierpinski triangle.}
	\label{fig:sierpinski}
\end{figure}

Formally, the discrete Sierpinski triangle is a set of points in $\Ints[2]$. Let $V=\set{(1,0), (0,1)}$ and define the sets $\S[0], \S[1], \ldots$ by the recursion
\begin{align}
	\begin{split}\label{eqn:Sstage}
    \S[0]   &=  \set{(0,0)}, \rm{\ and} \\
    \S[i+1] &=  \S[i] \cup (\S[i] + 2^iV),
    \end{split}
\end{align}
where $A + cB = \set{ \vm + c\vn \mid \vm \in A {\ \rm{and}\ } \vn \in B }$.  
The \emph{discrete Sierpinski triangle} is the set
\begin{equation}\label{eqn:S}
    \S = \bigcup_{i=0}^{\infinity} \S[i].
\end{equation}
We often  refer to $\S[i]$ as the $\ith$ stage of $\S$.
Note that $\S$ can also be defined as the nonzero residues modulo 2 of Pascal's triangle \cite{bondarenko1993a}.
It is also a numerically self-similar fractal \cite{sascrf}.
See Figure \ref{fig:sierpinski} for an illustration.

Using equation \ref{eqn:Sstage} it is easy to give a formula for the cardinality of the $i^{th}$ stage of $\S$.

\begin{observation}\label{obs:SnSize}
	For each $n\in\Nats$, $|\S[n]| = 3^n$.
\end{observation}

Then, using Observation \ref{obs:SnSize} and Equations \eqref{eqn:zetadim} and \eqref{eqn:Sstage}, we can easily calculate the $\zeta$-dimension of $\S$.
\begin{observation}\label{obs:DimZetaS}
	$\ZetaDim(\S) = \log_2{3}$.
\end{observation}

Winfree \cite{winfreephd} proved that $\S$ weakly self-assembles in the TAM, and Rothemund,
Papadakis, and Winfree \cite{asadst} achieved a molecular implementation of this self-assembly. More recently, Lathrop, Lutz, and Summers \cite{ssadst} proved that $\S$ cannot strictly self-assemble in the TAM.

\begin{theorem}[Lathrop, Lutz, and Summers \cite{ssadst}]
$\S$ cannot strictly self-assemble in the Tile Assembly Model.
\end{theorem}

%% file: limits.tex

\section{Limitations on Approximating the Sierpinski Triangle}\label{sec:limits}

In this section we present our first main theorem. We show that every set that strictly self-assembles disagrees with $\S$ on a set with $\zeta$-dimension at least that of $\S$. We then show that for subsets of $\S$, the limitation is even more severe.

We first establish a bound on the number of tile types needed for $\S[n]$ to strictly self-assemble. Our first lemma follows easily from the tree like structure of $\S$.

\begin{lemma}\label{lem:SnTree}
Let $T$ be a set of tile types and $\tau \in \Nats$. If $\alpha \in \assemblies$ and $\dom\alpha=\S[n]$ for some $n \in \Nats$, then for each $\vm \in \dom\alpha$ and $\vu \in U_2$,
$$
	\vm + \vu \in \dom\alpha
	\implies
	\alpha(\vm)(\vu)=\alpha(\vm+\vu)(-\vu) \textrm{ and } \gstr[\alpha(\vm)](\vu) \ge \tau.
$$
\begin{proof}
Assume the hypothesis with $T$, $\tau$, $\alpha$, and $n$ as witness. Let $\vm \in \dom\alpha$ and $\vu \in U_2$ such that $\vm + \vu \in \dom\alpha$.
It suffices to show that $\gstr[\alpha](\vm, \vm + \vu) \ge \tau$. Let $G_{\alpha}=(V,E,\beta)$ be the binding graph of $\alpha$. Note that since $\dom\alpha = \S[n]$, $(V,E)$ is a tree rooted at the origin and since $\alpha$ is $\tau$-stable, $\beta_{(V,E)} \ge \tau$. So, it suffices to show that $\beta((\vm, \vm+\vu)) \ge \tau$.

Since $(V,E)$ is a tree, and $\vm$ and $\vm + \vu$ are adjacent in $(V,E)$, either $\vm$ is on the path from the origin to $\vm + \vu$ or $\vm + \vu$ is on the path from the origin to $\vm$. Without loss of generality, assume $\vm$ is on the path from the origin to $\vm + \vu$ (otherwise, the the theorem holds for $\vm' = \vm+\vu$ and $\vu' = -\vu$).
Let $C=(C_0,C_1)$ be the unique cut of $G$ such that
\begin{align*}
	C_1 &= \set{
		\vn \in \dom\alpha
		\mid
		\text{$\vm + \vu$ is on a path in $G$ from the origin to $\vn$}
	}\text{, and}\\
	C_0 &= V\setminus C_1.
\end{align*}
Then, $\vm \in C_0$ and $\vm + \vu \in C_1$. Furthermore, since $(V,E)$ is a tree, $(\vm, \vm+\vu)\in E$ is the unique edge across $C$. But then, $\beta_C = \beta((\vm, \vm+\vu))$, and since $\beta_C \ge \beta_{(V,E)} \ge \tau$, $\beta((\vm, \vm+\vu)) \ge \tau$.
\end{proof}
\end{lemma}

\begin{lemma}\label{lem:SnTileComplexity}
If $\S[n]$ strictly self-assembles in a TAS $(T,\sigma,\tau)$, then $|T| \ge 2^{n+1}-1$.
\begin{proof}
Assume the hypothesis with $n \in \Nats$ and TAS $\calT = (T,\sigma,\tau)$ as witness.
Let $\alpha \in \terminals$.
If $n=0$ the lemma is trivially true, so assume $n > 1$.
Let $A = A_h \cup A_v$, where
$
	A_h = \set{(i,0) \mid 0 \le i < 2^n} \text{and } A_v = \set{(0,i) \mid 0 \le i < 2^n}.
$
Conceptually, $A_h$ (and $A_v$) represent the left (and right) boundary of $\S[n]$.
Let $T_A = \set{\alpha(\vm) \mid \vm \in A}$ be the set of all tile types placed at locations in $A$.
Clearly, $T_A \subseteq T$, so
it suffices to show that $|T_A| \ge 2^{n+1}-1$.
Suppose  $|T_A| < 2^{n+1}-1$.
By \eqref{eqn:Sstage} and $\alpha \in \terminals$, $\vm \in \dom\alpha$ for all $\vm \in A$.
Then, there exist a $\vm,\vn \in A$ such that $\vm \ne \vn$ and $\alpha(\vm) = \alpha(\vn)$.
Either $\vm,\vn \in A_h$, $\vm,\vn \in A_v$, or $\vm \in A_h \setminus \{\vzero\}$ and $\vn \in A_v \setminus \{\vzero\}$.
In each case we show that $\S[n]$ does not strictly self-assemble in $\calT$.

\begin{figure}[!ht]
	\centering
    	\subfloat[Case 1. The figure on the left shows an example $\alpha$ where $\vm=(2,0)$ and $\vn=(5,0)$. The tiles $\alpha(\vm)$ and $\alpha(\vn)$ are blue. The tile $\alpha(\vm+\ue)$ is red and the tile $\alpha(\vm + 2\ue)$ is orange. The figure on the right shows the $\beta$ for this $\alpha$.]
    	{\label{fig:SnTileComplexityCase1}
		\includegraphics{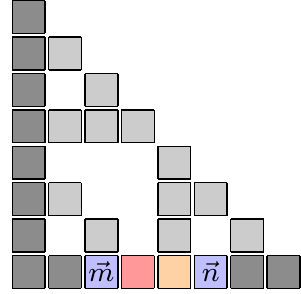}
		\quad
		\includegraphics{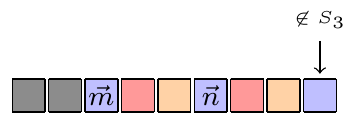}
	}
	\quad
	\subfloat[Case 3. The figure on the left shows an example $\alpha$ where $\vm=(2,0)$ and $\vn=(0,5)$.The tiles $\alpha(\vm)$ and $\alpha(\vn)$ are blue. The tile $\alpha(\vm+\uw)$ is red.  The figure on the right shows the $\beta$ for this $\alpha$.]
	{\label{fig:SnTileComplexityCase3}
		\includegraphics{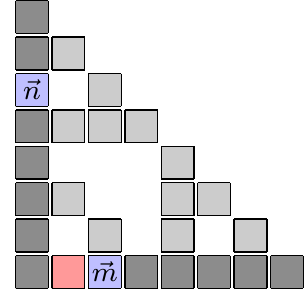}
		\quad
		\includegraphics{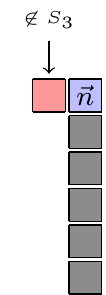}
	}
	\caption{Illustrating the proof of Lemma \ref{lem:SnTileComplexity} for $n=3$.}
	\label{fig:SnTileComplexity}
\end{figure}

\begin{itemize}
\item[]
Case 1.
Suppose $\vm,\vn \in A_h$.
Without loss of generality, let $\vm = (i,0)$ and $\vn = (j,0)$ where $0 \le i < j < 2^n$.
Let $\beta$ be the unique $T$-configuration such that for all $\vk=(k_1,k_2) \in \Nats$,
\begin{equation*}
	\beta(\vk) = \begin{cases}
			\uparrow & \tif\ k_1 > 2^n \tor\ k_2 \ne 0\\
			\alpha(\vk) & \tif\ k_1 < j\\
			\alpha(\vm + ((k_1-j) \bmod (j-i), 0)) & \totherwise.
		\end{cases}
\end{equation*}
See Figure \ref{fig:SnTileComplexityCase1} for an illustration.
By \eqref{eqn:Sstage} and $j > 0$, $\vn + \uw \in \dom\alpha$.
So, by Lemma \ref{lem:SnTree}, $\alpha(\vn)(\uw)=\alpha(\vn+\uw)(\ue)$ and
$\gstr[\alpha(\vn)](\uw) \ge \tau$.
But since $\alpha(\vn)=\alpha(\vm)$, $\alpha(\vm)(\uw)=\alpha(\vn+\uw)(\ue)$ and $\gstr[\alpha(\vm)](\uw) \ge \tau$.
So, $\beta \in \producibles$.
Then, there exists a $\gamma \in \terminals$ such that $\beta \sqsubseteq \gamma$ and since $\beta((2^n,0))\downarrow$, $\gamma((2^n,0))\downarrow$. But $(2^n,0) \not\in \S[n]$. So, $\S[n]$ does not strictly self-assemble in $\calT$.

\item[] Case 2. The case for $\vm,\vn \in A_v$ is similar to Case 1.

\item[]
Case 3.
Suppose $\vm \in A_h \setminus \{\vzero\}$ and $\vn \in A_v \setminus \{\vzero\}$.
Let $\vm=(i,0)$ and $\vn=(0,j)$, where $i,j\in\set{1,\ldots,2^n-1}$.
Let $\beta$ be the unique $T$-configuration such that for all $\vk=(k_1,k_2) \in \Nats$,
\begin{equation*}
	\beta(\vk) = \begin{cases}
			\alpha(\vk) & \tif\ k_1=0 \tand\ k_2 \le j \tor\  k_2=0 \tand\ k_1 \le i\\
			\alpha(\vm+\uw) & \tif\ k_1=-1 \tand\ k_2=j\\
			\uparrow & \totherwise.
		\end{cases}
\end{equation*}
See Figure \ref{fig:SnTileComplexityCase3} for an illustration.
By \eqref{eqn:Sstage} and $i > 0$, $\vm + \uw \in \dom\alpha$.
So, by Lemma \ref{lem:SnTree}, $\alpha(\vm)(\uw)=\alpha(\vm+\uw)(\ue)$ and $\gstr[\alpha(\vm)](\uw) \ge \tau$.
But since $\alpha(\vm)=\alpha(\vn)$, $\alpha(\vn)(\uw)=\alpha(\vm+\uw)(\ue)$ and $\gstr[\alpha(\vn)](\uw) \ge \tau$.
So, $\beta \in \producibles$.
Then, there exists a $\gamma \in \terminals$ such that $\beta \sqsubseteq \gamma$ and since $\beta((-1,j))\downarrow$, $\gamma((-1,j))\downarrow$.
But $(-1,j) \not\in \S[n]$. So, $\S[n]$ does not strictly self-assemble in $\calT$. \hfill \qedhere
\end{itemize}
\end{proof}
\end{lemma}

Even if we only require that $\S[n]$ appear somewhere
in the terminal assembly (not necessarily at the origin), we still have an exponential lower bound on the minimum number of tile types needed. To show this we use the \emph{ruler function} $\rho : \Ints[+] \rightarrow \Nats$ defined by the recurrence $\rho(2k+1) = 0$ and  $\rho(2k) = \rho(k) + 1$ for all $k \in \Nats$.
The value of $\rho(n)$ is the exponent of the largest power of 2 that divides $n$, or equivalently, $\rho(n)$ is the number of 0's lying to the right of the rightmost 1 in the binary representation of $n$ \cite{concrete}.
Now, for each $i \in \Nats$, the width of the longest horizontal bar rooted at $(0,i)$ and the height of the tallest vertical bar rooted at $(i,0)$ in $\S$ is $2^{\rho(i)}-1$ \cite{ssadst}.

\begin{lemma}\label{lem:SnAnywhereTileComplexity}
Let $n \in \Nats$ and $\vm \in \Ints[2]$. If $\calT = (T,\sigma,\tau)$ is a TAS such that for every $\alpha \in \terminals$, $\dom\alpha \cap (\vm + \set{0,\ldots,2^n-1}^2) = \vm + \S[n]$, then $|T| \ge 2^n-2$.
\begin{proof}
Assume the hypothesis with $n \in \Nats$, $\vm=(m_1,m_2)\in\Ints[2]$, and TAS $\calT=(T,\sigma,\tau)$ as witness.
Suppose $|T| < 2^n-2$. Then we can construct the TAS $\calT'=(T',\sigma',\tau)$ as follows.
\begin{figure}[!b]
	\centering
	\subfloat[The assembly $\alpha$ for some arbitrary $\vm$. Tile types added to $T'$ in step (1) are blue.]
	{
		\hspace{10pt}
		\includegraphics{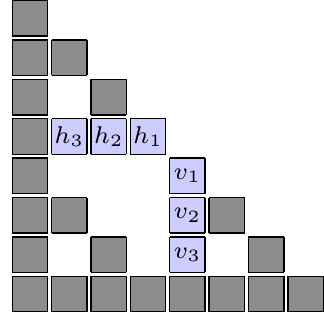}
		\hspace{10pt}
	}
	\quad
	\subfloat[The assembly $\alpha'$ such that $\dom\alpha'={\S[n]}$. Tile types added to $T'$ in steps (2) \& (3) are  gray.]
	{
		\hspace{10pt}
		\includegraphics{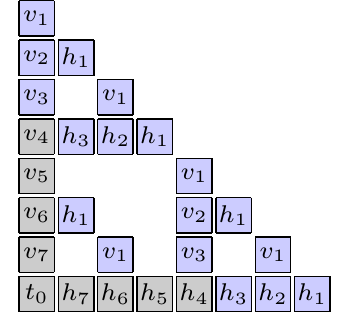}
		\hspace{10pt}
	}
	\caption{Illustrating the proof of Lemma \ref{lem:SnAnywhereTileComplexity} for $n=3$.}
	\label{fig:SnAnywhereTileComplexity}
\end{figure}
\begin{itemize}
\item[(1)] For each $0 < i < 2^{n-1}$, there are tile types $v_i, h_i \in T'$ such that
	\begin{align*}
		v_i = \alpha(\vm + (2^{n-1}, i)) \text{ and } h_i = \alpha(\vm + (i, 2^{n-1})).
	\end{align*}
	
\item[(2)] For each $2^{n-1} \le i < 2^n$, there are a tile types $v_i, h_i \in T'$ such that
	\begin{align*}
			 v_i(\un) &= \begin{cases}
					v_{i-1}(\us) &\tif\ i = 2^{n-1}\\
					(i-1, \tau) &\tif\ i > 2^{n-1}
					\end{cases}
		& &
		       h_i(\ue) = \begin{cases}
				h_{i-1}(\uw) &\tif\ i = 2^{n-1}\\
				(i-1, \tau) &\tif\ i > 2^{n-1}
			\end{cases}\\
             v_i(\ue) &= (h_{2^{\rho(i)}-1}(\uw), \tau) && h_i(\un) = (v_{2^{\rho(i)}-1}(\us), \tau)
    \end{align*}
    \begin{align*}
             v_i(\us) &= h_i(\uw) = (i, \tau) &&  v_i(\uw) = h_i(\us) = (0, 0).
    \end{align*}
   			
\item[(3)] There is a tile type $t_0 \in T'$ such that $\sigma'=t_0$, $t_0(\un) =t_0(\ue)=(2^n-1,\tau) \text{ and } t_0(\us)=t_0(\uw)=(nil,0)$.
\end{itemize}
Each tile type added to $T'$ in step (1) is also a tile type in $T$, so we add at most $2^n-3$ tile types to $T'$ in step (1).
We add a total of $2^n$ tile types to $T'$ in step (2) and we add 1 tile type to $T'$ in step (3). Thus, $|T'| \le 2^{n+1} - 2$.
But, by using \eqref{eqn:Sstage} and the ruler function properties, it is easy to verify that $\S[n]$ strictly self-assembles in $\calT'$. This contradicts Lemma \ref{lem:SnTileComplexity}. Thus, no such TAS exists. 
\end{proof}
\end{lemma}

We now have the necessary machinery to prove our first main theorem which says that every set that strictly self-assembles disagrees with $\S$ on a set with fractal dimension at least that of $\S$. Hence, $\S$ cannot even be approximated closely with strict self-assembly.

\begin{theorem}\label{thm:close}
If $X \subseteq \Ints[2]$ strictly self-assembles, then $\ZetaDim(X \Delta\, \S) \ge \ZetaDim(\S)$.
\begin{proof}
Assume the hypothesis with $X\subseteq\Ints[2]$ and TAS $\calT=(T,\sigma,\tau)$ as witness. Let $$V = \set{(0,0),(0,1),(1,0)}$$ and $n=\ceil{\log_2{(|T|+2)}+1}$. Since $X$ strictly self-assembles in $\calT$, for every $\alpha \in \terminals$, $X = \dom\alpha$. Let $d:\Ints[2]\times\Nats\rightarrow\Nats$ where
\begin{equation}\label{eqn:defofd}
	d(\vm,k) = \left|(X \cap (\vm+\{0,\ldots,2^{n+k}-1\}^2))\, \Delta\, (\vm + \S[n+k])\right|
\end{equation}
for all $\vm \in \Ints[2]$ and $k \in \Nats$.
Then, by \eqref{eqn:Sstage},
$
	d(\vm,k) \ge \sum_{\vv \in V} d(\vm+ 2^{n+k-1}\vv, k-1).
$
Since $|T|<2^n-2$, by Lemma \ref{lem:SnAnywhereTileComplexity}, for all $\vm\in\Ints[2]$, $X \cap (\vm+\set{0,\ldots,2^n-1}^2) \ne \vm+\S[n]$. So, for all $\vm\in\Ints[2]$, $d(\vm, 0) \ge 1$. 
Then, the recurrence solves to 
\begin{equation}\label{eqn:sizeofd}
        d(\vm,k) \ge 3^k
\end{equation}
 for all $\vm\in\Ints[2]$.
So,
\begin{align*}
    \ZetaDim(X \Delta\,\S) &\stackrel{\eqref{eqn:zetadim}}{=}  \limsup_{n \rightarrow \infty} \frac{\log_2{|X_{[0,2^n)}|}}{n}\\
    				&= \limsup_{k \rightarrow \infinity} \frac{\log_2{d(\vec{0},k)}}{n+k} &\text{by Equation \eqref{eqn:defofd}}\\
                                     &\ge \limsup_{k \rightarrow \infinity} \frac{\log_2{3^k}}{n+k} &\text{by Equation \eqref{eqn:sizeofd}}\\
 	                            &= \log_2{3}.
\end{align*}
But, by Observation \ref{obs:DimZetaS}, $\ZetaDim(\S)=\log_2{3}$.
\end{proof}
\end{theorem}

To gain further insight, we now consider the strict self-assembly of subsets of $\S$, and show that here the limitation is even more severe.
We first give an upper bound on the number of tiles located within a given distance of the seed tile in any strict self-assembly of a subset of $\S$.
We use the a theorem from \cite{ssadst} that for any structure to strictly self-assemble, the number of tile types used is at least the finite-tree depth of the structure.

\begin{theorem}[Lathrop, Lutz, and Summers \cite{ssadst}]\label{thm:T-ge-depth}
If $X\subseteq \Ints[2]$ strictly self-assembles in a TAS $(T,\sigma,\tau)$, then $|T|\ge\ftdepth{\dom\sigma}(G^{\#}_{X})$.
\end{theorem}

It easily follows that in any strict self-assembly of a subset of $\S$, not to many tiles can be placed far from the boundary.

\begin{corollary}\label{cor:CloseToBoundary}
If $X \subseteq \S$ strictly self-assembles in a TAS $(T,\sigma,\tau)$, then for all $\vec{m}=(m_1,m_2) \in \Ints[2]$ such that $m_1 \ge |T|$ and $m_2 \ge |T|$, $\vec{m} \not\in X$.
\end{corollary}

\begin{lemma}\label{lem:SizeOfSubset}
If $X \subseteq \S$ strictly self-assembles in a TAS $(T,\sigma,\tau)$, then for every $n \in \Nats$, $|X_{[0,n]}| \le 2|T|(n+1)$.
\begin{proof}
Assume the hypothesis with $X\subseteq\Ints[2]$ and $\calT=(T,\sigma,\tau)$ as witness.
Let $\alpha \in \terminals$ and let $n \in \Nats$.
If $n \le |T|$ the theorem is trivially true, so assume $n > |T|$.
Let $A = \set{0,\ldots,|T|-1}^2$, $B=\set{0,\ldots,|T|-1}\times\set{|T|,\ldots,n}$, $C=\set{|T|,\ldots,n}\times\set{0,\ldots,|T|-1}$, and $D=\set{|T|,\ldots,n}^2$.
It is clear that $A, B, C, D$ is a partition of $\set{0,\ldots,n}^2$.
Then,
\begin{align*}
    |X_{[0,n]}| &= |\set{0,\ldots,n}^2 \cap \dom{\alpha}| & \text{since $X \subseteq \Nats[2]$}\\
				&= |A \cap \dom{\alpha}| + |B \cap \dom{\alpha}| + |C \cap \dom{\alpha}| & \text{by Corollary \ref{cor:CloseToBoundary}}\\
				&\le |T|^2 + 2|T|(n-|T|+1)\\
				&\le 2|T|(n+1). & \hfill \qedhere
\end{align*}
\end{proof}
\begin{figure}[!ht]
	\centering
		\includegraphics{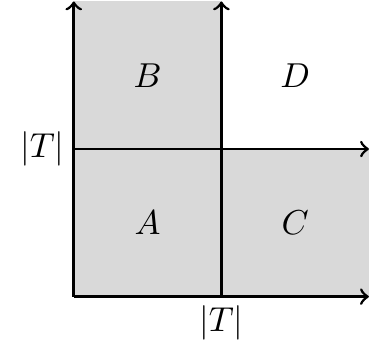}
	\caption{Illustrating the partition used in the proof of Lemma \ref{lem:SizeOfSubset}}
	\label{fig:SizeOfSubset}
\end{figure}
\end{lemma}

\begin{theorem}\label{thm:Subset}
If $X \subseteq \S$ strictly self-assembles, then $\ZetaDim(X) \in \set{0,1}$.
\begin{proof}
Assume the hypothesis with $X\subseteq\S$ and TAS $(T,\sigma,\tau)$ as witness.
By Lemma \ref{lem:SizeOfSubset}, $|X_{[0,n]}| \le 2|T|(n+1)$. Then, $\ZetaDim(X) \le 1$.
But, the binding graph of any $\alpha\in\assemblies$ must be connected and any infinite connected structure has $\zeta$-dimension at least 1, it follows that either $\ZetaDim(X)=1$ or $X$ is finite, in which case $X$ has $\zeta$-dimension 0. So, $\ZetaDim(X)\in\set{0,1}$.
\end{proof}
\end{theorem}

Note that boundary of $\S$ is a subset of $\S$ that strictly self-assembles and has $\zeta$-dimension 1. A single tile placed at the origin is a subset of $\S$ that strictly self-assembles and has $\zeta$-dimension 0. Hence, Theorem \ref{thm:Subset} is trivially tight.

%% file: cd.tex

\section{Conditional Determinism}\label{sec:cd}

\begin{figure}[!ht]
	\centering

    	\subfloat[The tile set $T_B$.]
    	{\label{fig:BlockingTiles}
    		\includegraphics{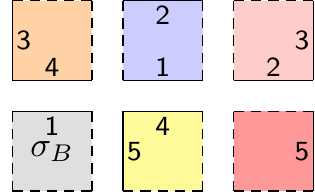}
    	}
    	\quad
    	\subfloat[$\res(\valpha)$. The site of the blocking is marked with a red X.]
    	{\label{fig:BlockingResult}
    		\hspace{3em}
    		\includegraphics{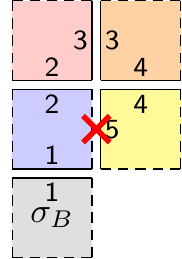}
    		\hspace{3em}
    	}
    	\quad
    	\subfloat[$\valpha$ fails condition (2) of local determinism for the location $(0,1)$.]
    	{\label{fig:BlockingRipped}
    		\hspace{3em}
    		\includegraphics{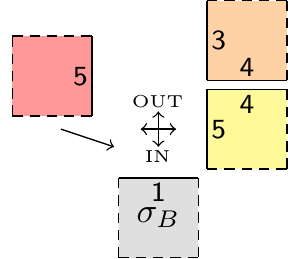}
    		\hspace{3em}
    	}
	\caption{The TAS $\calT_B=(T_B,\sigma_B,1)$ which uses a blocking technique.}
	\label{fig:Blocking}
\end{figure}

The method of local determinism introduced by Soloveichik and Winfree \cite{csas} is a common technique for showing that a TAS is directed.  However, there exists very natural constructions that are directed but not locally deterministic. Consider the TAS $\calT_B=(T_B,\sigma_B,1)$ of Figure \ref{fig:Blocking}.
Clearly, there is only one assembly sequence $\valpha$ in $\calT_B$ such that $\res\valpha$ is terminal. Hence, $\calT_B$ is directed. However, $\valpha$ fails condition (2) of local determinism at the location $(0,1)$. The culprit is the blocking technique used by this TAS which is marked by a red X in Figure \ref{fig:BlockingResult}. Since $\valpha$ is the only possible locally deterministic assembly sequence in $\calT_B$, then $\calT_B$ is not a locally deterministic TAS.  Thus, new techniques are needed to show this TAS is directed.

In this section we give sufficient conditions for proving such a TAS is directed. First, we introduce some new notation. For $\vm,\vn\in\Ints[2]$, if $\vm \prec_{\valpha} \vn$ for every assembly sequence $\valpha$ in a TAS $\calT$, then we say \emph{$\vm$ precedes $\vn$ in $\calT$}, and we write $\vm \prec_{\calT} \vn$. For each $\vm \in \Ints[2]$, we define the set
\begin{align*}
	\depsides(\vm) &= \set{\vu \in U \mid \vm \prec_{\calT} \vm + \vu}.
\end{align*}
Now, let $\calT$ be a TAS, $\valpha$ an assembly sequence in $\calT$, and $\alpha=\res\valpha$.
Then, $\valpha$ is \emph{conditionally deterministic} if the following three conditions hold.
\begin{itemize}
\item[(1)] For all $\vm \in \dom\alpha \setminus \dom\alpha_0$,
	 $
		\sum_{\vu\in\insides(\vm)} \gstr[\alpha_{i_{\valpha}(\vm)}](\vm,\vm+\vu) = \tau
	 $.
\item[(2)] For all $\vm \in \dom\alpha\setminus\dom\alpha_0$ and all $t \in T \setminus \set{\alpha(\vm)}$,
	$$
		\vm \not\in \tfrontier{t} (\alpha \upharpoonright (\dom\alpha \setminus (\set{\vm} \cup (\vm + (\outsides(\vm) \cup \depsides(\vm)))))).
	$$
\item[(3)] $\frontier \alpha = \emptyset$.
\end{itemize}
Note that conditions (1) and (3) are the same as in the definition of local determinism.
Conceptually, (1) requires that each tile added in $\valpha$ \dquote{just barely} binds to the existing assembly; (2) holds when the tiles at $\vm$ and $\vm + \outsides(\vm) + \depsides(\vm)$ are removed from $\alpha$, no other tile type can attach to the assembly at location $\vm$; and (3) requires that $\alpha$ is terminal.
A TAS is \emph{conditionally deterministic} if it has a conditionally deterministic assembly sequence.

Our first theorem shows that conditional determinism is a weaker notion than local determinism.

\begin{theorem}\label{thm:LDimpliesCD}
Every locally deterministic TAS is conditionally deterministic.
\begin{proof}
Let $\calT$ be a locally deterministic TAS with $\valpha=(\alpha_i \mid 0 \le i < k)$ as witness.
Let $\alpha=\res(\valpha)$.
It suffices to show that $\valpha$ is a conditionally deterministic assembly sequence.
Since conditions (1) and (3) in the definitions of both local determinism and conditional determinism are the same, it suffices to show that condition (2) in the definition of conditional determism holds for $\valpha$.
Since $\valpha$ is locally deterministic, by condition (2) of local determinism, for all $\vm \in \dom\alpha \setminus \dom\alpha_0$ and all $t \in T \setminus \set{\alpha(\vm)}$,
$$\vm \not\in \tfrontier{t} (\alpha \upharpoonright (\dom\alpha \setminus (\set{\vm} \cup (\vm + \outsides(\vm))))).$$
Then, since $\outsides(\vm) \subseteq \outsides(\vm) \cup \depsides(\vm)$, it follows that for all $\vm \in \dom\alpha \setminus \dom\alpha_0$ and all $t \in T \setminus \set{\alpha(\vm)}$,
 $$\vm \not\in \tfrontier{t} (\alpha \upharpoonright (\dom\alpha \setminus (\set{\vm} \cup (\vm + (\outsides(\vm) \cup \depsides(\vm)))))).$$
Hence, $\valpha$ is a conditionally deterministic assembly sequence in $\calT$.
\end{proof}
\end{theorem}

We now show that although conditional determinism is weaker than local determinism, it is strong enough to show a TAS is directed. 

\begin{theorem}\label{thm:CDimpliesDirected}
Every conditionally deterministic TAS is directed.
\begin{proof}
Our proof is similar to the proof in \cite{csas} that every locally deterministic TAS is directed.
Let $\calT=(T,\sigma,\tau)$ be a conditionally deterministic TAS with $\valpha = (\alpha_i \mid 0 \le i < k)$ as witness.
Let $\alpha=\res(\valpha)$ and note that $\alpha \in \terminals$.
To see that $\calT$ is directed, it suffices to show that for all $\beta \in \terminals$, $\beta \sqsubseteq \alpha$.

Let $\beta \in \terminals$. Then, there is an assembly sequence $\vbeta = (\beta_j \mid 0 \le j < l)$ in $\calT$ such that $\beta_0=\sigma$ and $\beta = \res(\vbeta)$. To see that $\beta \sqsubseteq \alpha$, it suffices to show that for each $0 \le j < k$, the following conditions hold:
\begin{itemize}
\item[(1)] $\insides[\vbeta](\dom\beta_{j+1} \setminus \dom\beta_j) = \insides[\valpha](\dom\beta_{j+1} \setminus \dom\beta_j)$, and
\item[(2)] $\beta(\dom\beta_{j+1} \setminus \dom\beta_j) = \alpha(\dom\beta_{j+1} \setminus \dom\beta_j)$.
\end{itemize}

Suppose there exists a $0 \le j < k$ such that either condition (1) or condition (2) fails. Let $i$ be the smallest such $j$. To prove the theorem, it suffices to show no such $i$ exists. Let $\vb_i = \dom\beta_{i+1} \setminus \dom\beta_i$.
Consider any $\vu \in \insides[\vbeta](\vb_i)$.
It is clear that $-\vu \in \outsides[\vbeta](\vb_i + \vu)$, so $-\vu \not\in \insides[\vbeta](\vb_i + \vu)$.
Either $\vb_i + \vu \in \dom\sigma$ or there exists an $h < i$ such that $\vb_h = \vb_i + \vu$.\\
\begin{itemize}
\item[] Case 1.
Suppose $\vb_i + \vu \in \dom\sigma$.
Then, since both $\valpha$ and $\vbeta$ are assembly sequences in $\calT$, $\insides[\vbeta](\vb_i + \vu) = \insides[\valpha](\vb_i + \vu) = \emptyset$.
Then, $-\vu \not\in \insides[\valpha](\vb_i+\vu)$.
So, $\vu \not\in \outsides[\valpha](\vb_i)$.
Also, $i_{\vbeta}(\vb_i + \vu)=0$, so $\vb_i \not\prec_{\calT} \vb_i + \vu$.
Then, $\vu \not\in \depsides(\vb_i)$.
\item[] Case 2.
Suppose there exists an $h < i$ such that $\vb_h = \vb_i + \vu$.
Then, by condition (1), $\insides[\vbeta](\vb_i + \vu) = \insides[\valpha](\vb_i + \vu)$.
Then, $-\vu \not\in \insides[\valpha](\vb_i+\vu)$.
So, $\vu \not\in \outsides[\valpha](\vb_i)$.
Also, $h \prec_{\vbeta} i$, so $\vb_i \not\prec_{\calT} \vb_i + \vu$.
Then, $\vu \not\in \depsides(\vb_i)$.
\end{itemize}
In either case,
\begin{equation*}
	\insides[\vbeta](\vb_i) \cap (\outsides[\valpha](\vb_i) \cup \depsides(\vb_i)) = \emptyset. \tag{i}
\end{equation*}
Since for all $\vm \in \dom\beta_i, \beta_i(\vm)=\alpha(\vm)$, then for all $\vu \in U_2$,
\begin{equation*}
	\gstr[\beta_{i+1}](\vb_i,\vb_i+\vu) \le \gstr[\alpha](\vb_i,\vb_i+\vu). \tag{ii}
\end{equation*}
Then, by (i) and (ii),
\begin{equation*}
	\Sigma_{\vu \in \insides[\vbeta](\vb_i)} \gstr[\beta_{i+1}](\vb_i, \vb_i + \vu)
	\le
	\Sigma_{\vu \in U_2\setminus(\outsides[\valpha](\vb_i)\cup\depsides(\vb_i))} \gstr[\alpha](\vb_i, \vb_i + \vu).
\end{equation*}
But, by property (2) of conditional determinism, the only type of tile that can attach to $\beta_i$ at location $\vb_i$ is $\alpha(\vb_i)$. Thus, $\beta(\vb_i) = \alpha(\vb_i)$.

So it must be the case that $\insides[\vbeta](\vb_i) \ne \insides[\valpha](\vb_i)$.
By property (1) of conditional determinism, there must be some $\vu \in \insides[\vbeta](\vb_i) \setminus \insides[\valpha](\vb_i)$.
Since $\vu \in \insides[\vbeta](\vb_i)$, $\vb_i + \vu \in \dom\beta_i$, so $\beta(\vb_i + \vu) = \alpha(\vb_i + \vu)$. We've already established that $\beta(\vb_i) = \alpha(\vb_i)$. So, By property (2) of conditional determinism, it must be the case that $i_{\valpha}(\vb_i + \vu) > i_{\valpha}(\vb_i)$. So, $\vu \not\in \insides[\valpha](\vb_i)$. But then $-\vu \in \insides[\valpha](\vb_i + \vu)$, and so $\vu \in \outsides[\valpha](\vb_i)$. But, by (i), this is impossible. Therefore, no such i exists.
\end{proof}
\end{theorem} 

It is now a straightforward task to show that the TAS of Figure \ref{fig:Blocking} is directed.

%% file: laced.tex

\newcommand{\TASL}{\calT_{\L}}
\newcommand{\TL}{T_{\L}}
\renewcommand*{\L}[1][]{\ensuremath{\if\blank{#1}{\mathrm{\bf{L}}}\else{\mathrm{\bf{L}}_{#1}}\fi}}
\newcommand{\Caps}[1][]{\ensuremath{C_{#1}}}
\newcommand{\Counters}[1][]{\ensuremath{N_{#1}}}
\newcommand{\Tests}[1][]{\ensuremath{T_{#1}}}

\section{Fibering the Sierpinski Triangle in Place}\label{sec:laced}

\begin{figure}[h!t!]
	
	\centering

	\subfloat{
		\centering
		\includegraphics[width=5.5in]{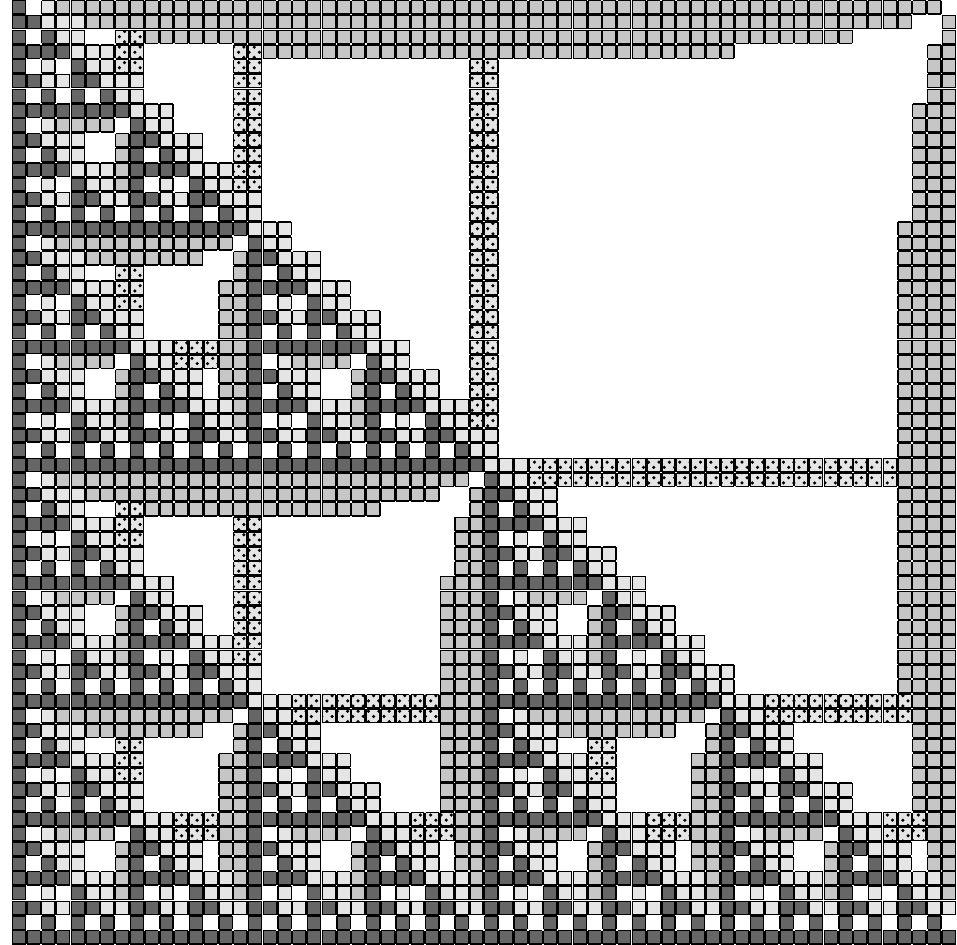}
	}
	\quad
	\begin{minipage}[b]{3.5in}
		\centering
		\includegraphics{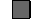}{\ \scriptsize $\S$}
		\includegraphics{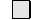}{\ \scriptsize Cap Fiber}
		\includegraphics{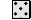}{\ \scriptsize Test Fiber}
		\includegraphics{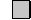}{\ \scriptsize Counter Fiber}
	\end{minipage}

	\caption{Stage 6 of the laced Sierpinski triangle.}

	\label{fig:LacedSierpinski}

\end{figure}

In this section we present our second main theorem. We construct a TAS in which a superset of $\S$ with the same $\zeta$-dimension strictly self-assembles. Thus, our first main theorem is tight, even when restricted to supersets of $\S$. To prove this we define a new fractal, the \emph{laced Sierpinski triangle}, denoted $\L$. We show that $\S \subseteq \L$,  $\ZetaDim(\L\Delta\,\S)=\ZetaDim(\S)$, and that $\L$ strictly self-assembles in the Tile Assembly Model.

Formally, the laced Sierpinski triangle is a set of points in $\Ints[2]$.
Our goal is to define the sets $\L[0], \L[1], \ldots$ such that each $\L[i]$ is the $\ith$ stage in our construction of $\L$. We will break each $\L[i]$ up into disjoint subsets representing the different \dquote{types} of fibers added to $\S$ that allow $\L$ to strictly self-assemble.
Let $V = \set{(0,1),(1,0)}$, $W = \set{(0,0)} \cup V$, and 
\begin{equation*}
X = \set{(0,2),(2,0),(-1,1),(1,-1)} \cup W.
\end{equation*}
Then, we define the sets $\Caps[0], \Caps[1], \ldots$ by
	\begin{equation}\label{eqn:Lcaps}
		\Caps[i] =
		\begin{cases}
			\emptyset &\text{if $i < 2$}\\
			2^{i-1}(1,1) + W &\text{if $i=2$}\\
			(2^{i-1}(1,1) + X)\, \cup \bigcup\limits_{w \in W} (2^{i-1}w + \Caps[i-1]) &\text{otherwise}.
		\end{cases}
	\end{equation}
Intuitively, each $\Caps[i]$ is the set of \emph{cap fibers} in $\L[i]$. 
For each $i \in \Nats$, let
	\begin{equation}\label{eqn:Lcountersinc}
		\Phi_i = \bigcup\limits_{j=1}^{i-2} \bigcup\limits_{k=2^j+1}^{2^i-2^j} \set{(2^i-j,k),(k,2^i-j)}.
	\end{equation}
Note that $\Phi_i = \emptyset$ for $i < 3$. Then, we define the sets $\Counters[0], \Counters[1], \ldots$ by
	\begin{equation}\label{eqn:Lcounters}
		\Counters[i] =	
		\begin{cases}
			\emptyset & \rm{if}\ i < 3\\
			\Phi_i \cup \Counters[i-1] \cup \bigcup\limits_{v \in V} (2^{i-1}v + (\Counters[i-1] - \Phi_{i-1})) & \rm{otherwise}.
		\end{cases}
	\end{equation}
Intuitively, $\Counters[i]$ is the set of \emph{counter fibers} that run along the top and right sides of the empty triangles that form in the negative space around the interior of $\S[i+1]$.
For each $i \in \Nats$, let
	\begin{equation}\label{eqn:Ltestsinc}
		\Psi_i = \bigcup\limits_{j=3}^{i-1} \bigcup\limits_{k=2^i-2^j+3}^{2^i-j} \set{(2^j-1,k),(2^j,k),(k,2^j-1),(k,2^j)}.
	\end{equation}
Note that $\Psi_i = \emptyset$ for $i < 4$.
Then, we define the sets $\Tests[0], \Tests[1], \ldots \subseteq \Ints[2]$ by
	\begin{equation}\label{eqn:Ltests}
		\Tests[i] =	
		\begin{cases}
			\emptyset & \rm{if}\ i < 4\\
			\Psi_i \cup \Tests[i-1] \cup \bigcup\limits_{v \in V} (2^{i-1}v + (\Tests[i-1] - \Psi_{i-1})) & \rm{otherwise}.
		\end{cases}
	\end{equation}
Intuitively, $\Tests[i]$ is the set of \emph{test fibers} between the counter fibers and cap fibers in $\L[i]$.
Now, for each $i \in \Nats$, let
\begin{equation}\label{eqn:Lstage}
	\L[i] = \S[i] \cup \Caps[i] \cup \Counters[i] \cup \Tests[i].
\end{equation}
Then, the \emph{laced Sierpinski triangle} is the set
\begin{equation}\label{eqn:L}
	\L = \bigcup_{i=0}^{\infty} \L[i].
\end{equation}
We often refer to $\L[i]$ as the $\ith$ stage of $\L$. See Figure \ref{fig:LacedSierpinski} for an illustration.
From equation (\ref{eqn:Lstage}), it is clear that $\L$ is a superset of $\S$.

\begin{observation}\label{obs:SsubsetL}$\S \subseteq \L$.\end{observation}

We now show that the $\zeta$-dimension of $\L \Delta\,\S$ (hence also of $\L$) is the same as the $\zeta$-dimension of $\S$.

\begin{theorem}\label{thm:LZetaDim}
$\ZetaDim(\L \Delta\,\S)=\ZetaDim(\S)$.
\begin{proof}
Since $\S \subseteq \L$, it suffices to show that $\ZetaDim(\L\setminus\S)=\ZetaDim(\S)$.
By (\ref{eqn:Lstage}), for each $n \in \Nats$,
	$|(\L \setminus\S)_{[0,2^n)}| =  |\Caps[n]| + |\Counters[n]| + |\Tests[n]|.$
By (\ref{eqn:Lcaps}),
\begin{equation*}
	|\Caps[n]| =
		\begin{cases}
		0 & \rm{if}\ n<2\\
		3 & \rm{if}\ n=2\\
		3|\Caps[n-1]|+7 & \rm{otherwise}.
		\end{cases}
\end{equation*}
Solving this recurrence for $n \ge 2$ gives $|\Caps[n]| = 6.5 \cdot 3^{n-2} - 3.5$.
By (\ref{eqn:Lcountersinc}), for $n \ge 2$, $|\Phi_n| = 2^{n+1}(n-3)+8$.
Then, by (\ref{eqn:Lcounters}),
\begin{equation*}
	|\Counters[n]| =
		\begin{cases}
		0 & \rm{if}\ n < 3\\
		8 & \rm{if}\ n = 3\\
		3|\Counters[n-1]| + |\Phi_n| - 2 |\Phi_{n-1}| & \rm{otherwise}.
		\end{cases}
\end{equation*}
Solving this recurrence for $n \ge 2$ gives $|\Counters[n]| = 4 \cdot 3^{n-1} - 2^{n+2} + 4$.
By (\ref{eqn:Ltestsinc}), for $n \ge 3$, $|\Psi_n| = 2^{n+2} - 2n^2 - 6n + 4$.
Then, by (\ref{eqn:Ltests}),
\begin{equation*}
		|\Tests[n]| =
			\begin{cases}
			0 & \rm{if}\ n \le 3\\
			3|\Tests[n-1]| + |\Psi_n| - 2 |\Psi_{n-1}| & \rm{otherwise}.
		\end{cases}
\end{equation*}
Solving this recurrence for $n \ge 3$ gives $|\Tests[n]| = 3^{n-3}(n + 16.5) - 3n^2 - 9n + 34.5$.
Then,
\begin{align*}
	\ZetaDim(\L \Delta\,\S)
	&= \limsup_{n \rightarrow \infinity} \frac{\log_2 ( |\Caps[n]| + |\Counters[n]| + |\Tests[n]| )}{n} \\
	&= \limsup_{n \rightarrow \infinity} \frac{\log_2 ( 3^{n-3}(n+72) - 2^{n+2} -3n(n+3) + 35  )}{n}\\
    &= \ZetaDim(\S) \text{\ by Observation \ref{obs:DimZetaS}}. & \hfill \qedhere
\end{align*}
\end{proof}
\end{theorem}

It remains to show that $\L$ strictly self-assembles.  Our proof is constructive in that we exhibit a TAS in which $\L$ strictly self-assembles. We begin by explaining the general techniques used to fiber $\S$ in place, i.e., strictly self-assembly a superset of $\S$ without disturbing the set $\S$, and then delve into the details of a TAS implementing those techniques.\\

\begin{figure}[!ht]
	\centering
	\includegraphics[width=5in]{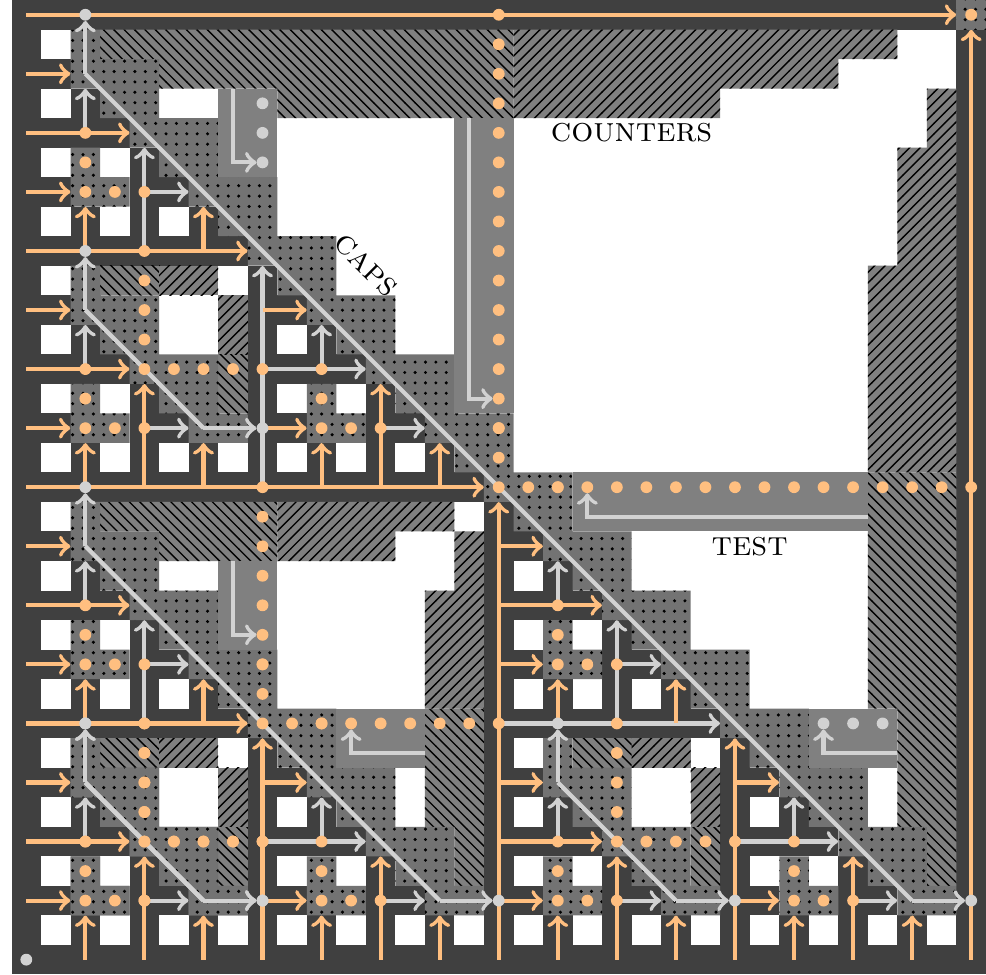}
	\caption{Fibering $\S$ in place.}
	\label{fig:LacedAssembly}
\end{figure}

Conceptually, the communication fibers added to $\S$ enable a superset of $\S_{i+1}$ to strictly self-assemble  when given a superset of $\S_{i}$ as input.  By \eqref{eqn:Sstage}, $\S_{i+1}$ can be constructed by placing a copy of $\S_{i}$ on top and to the right of itself.  This is achieved by copying the left boundary of $\S_{i}$ to the right of $\S_{i}$, and the bottom boundary of $\S_{i}$ to the top of $\S_{i}$.
These communication fibers are divided into three functional groups.
To ensure that the newly added bars are of the  proper length, \emph{counter} fibers control their attachment. The counter fibers increment until  they have reached same height as the middle point of the largest diagonal in $\S_i$, and then decrement to zero.
To know where the middle point is, the counter fibers initiate the attachment of \emph{test fibers} which grow back to $\S_i$, test whether the middle point is reached, and return the result to the counters.
However, if $\S_i$ has not yet fully attached, the test fibers will read from the wrong location. Nor can the test fibers wait until $\S_i$ has completed attaching before returning to the counters, because the test fibers would have to know where to wait!
The solution to this is the \emph{diagonal cap} fibers that attach along the largest diagonal in $\S$ on the side opposite the seed. The purpose of the diagonal cap fibers is to force the necessary part of $\S_i$ to complete attaching before its middle is read by the test fibers. Then, a blocking technique can be used for the test fibers. The bottom row of the test fibers runs from the counters until blocked by the cap fibers. This attachment forms a path on which information can propagate from the diagonals back to the counters in a controlled manner. 
This is achieved by the \emph{diagonal cap} fibers that attach along the largest diagonal in $\S$ on the side opposite the seed.
They force the necessary part of $\S_i$ to complete attaching before the counters for $\S_{i+1}$ can begin to attach.
Then, a blocking technique is used for the test fibers. The bottom row of the test fibers runs from the counters until blocked by the cap fibers. This attachment forms a path on which information can propagate from the diagonals back to the counters in a controlled manner. See Figure \ref{fig:LacedAssembly} for an illustration.
We now describe how the self-assembly determines the center of $\S_i$. A location is at the center of $\S_i$ when it sits directly above the left boundary of the $\S_{i-1}$ structure on the right part of $\S_i$ and directly to the right of the bottom boundary of the $\S_{i-1}$ structure on the top part of $\S_i$.
This is computed in our construction by assigning to each bar of $\S$ a boolean value that is true (represented in Figure \ref{fig:LacedAssembly} by orange) only if it meets the criteria above.
Every new bar that attaches to an existing bar will carry a true value unless it is the unique bar that attaches at the halfway point.
Then, when two true bars meet, it is always at a location in the middle of the largest diagonal of some stage of $\S$.
When this is the case, it is noted by the diagonal cap fibers so it can be passed to the test fibers.
Note that every bar that attaches on the boundary has a true value.

\begin{figure}[!ht]
	\centering
    \captionsetup{margin=5em}
    \subfloat
    {
        \includegraphics[width=5.75in]{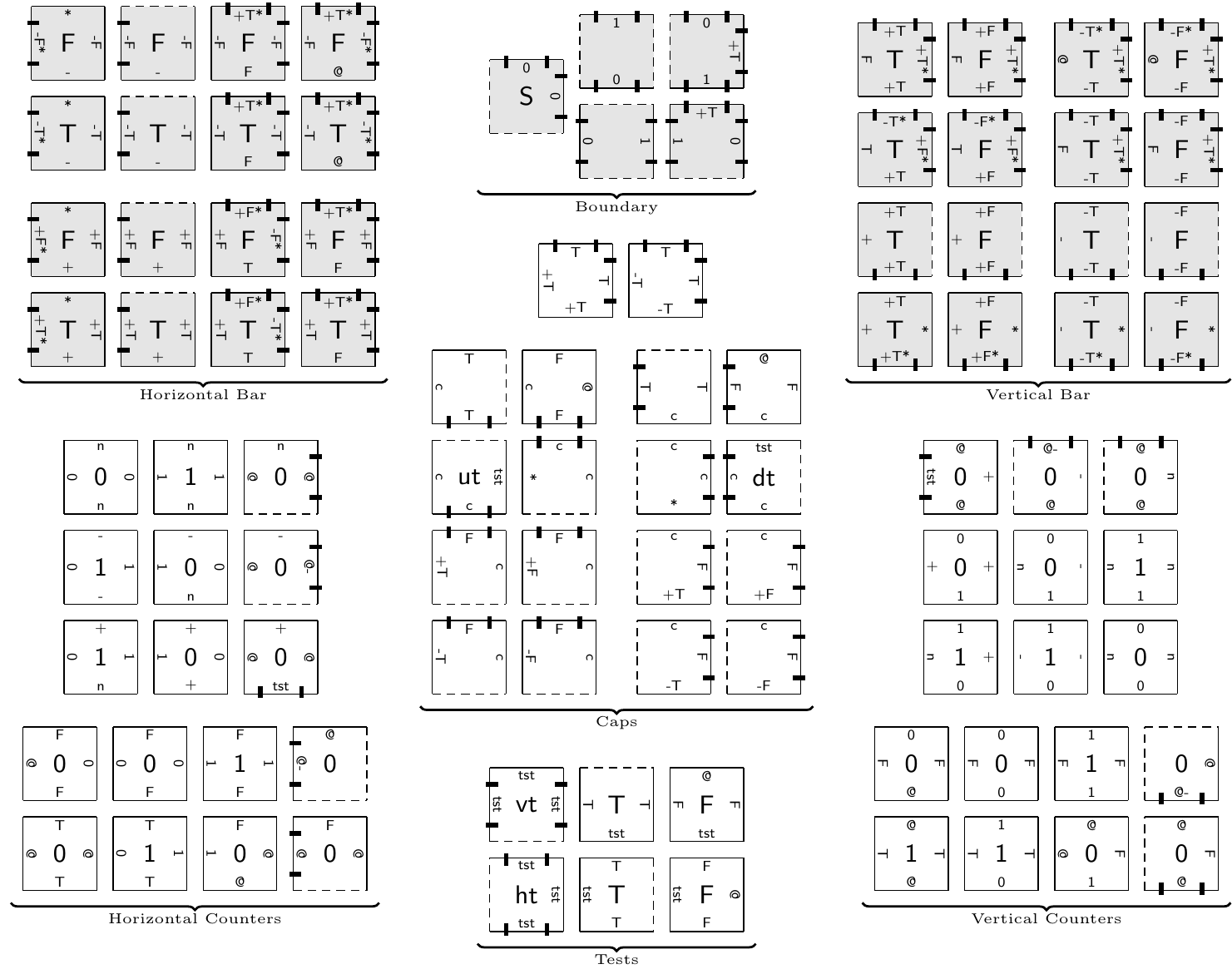}
    }
	\caption{The tile set $\TL$ of the TAS $\TASL$.}
	\label{fig:LacedTiles}
\end{figure}

We now construct a TAS $\TASL$ that implements the techniques described. 
Let $\TASL=(\TL, \sigma_{\L}, 2)$ be a TAS such that the set
$\TL$ has ninety-five tile types as illustrated in Figure \ref{fig:LacedTiles}, and $\sigma_{\L}$ is a tile of type $\scrib{S}$ from Figure \ref{fig:LacedTiles}.

\begin{figure}[!ht]
	\centering
        \includegraphics[width=5.75in]{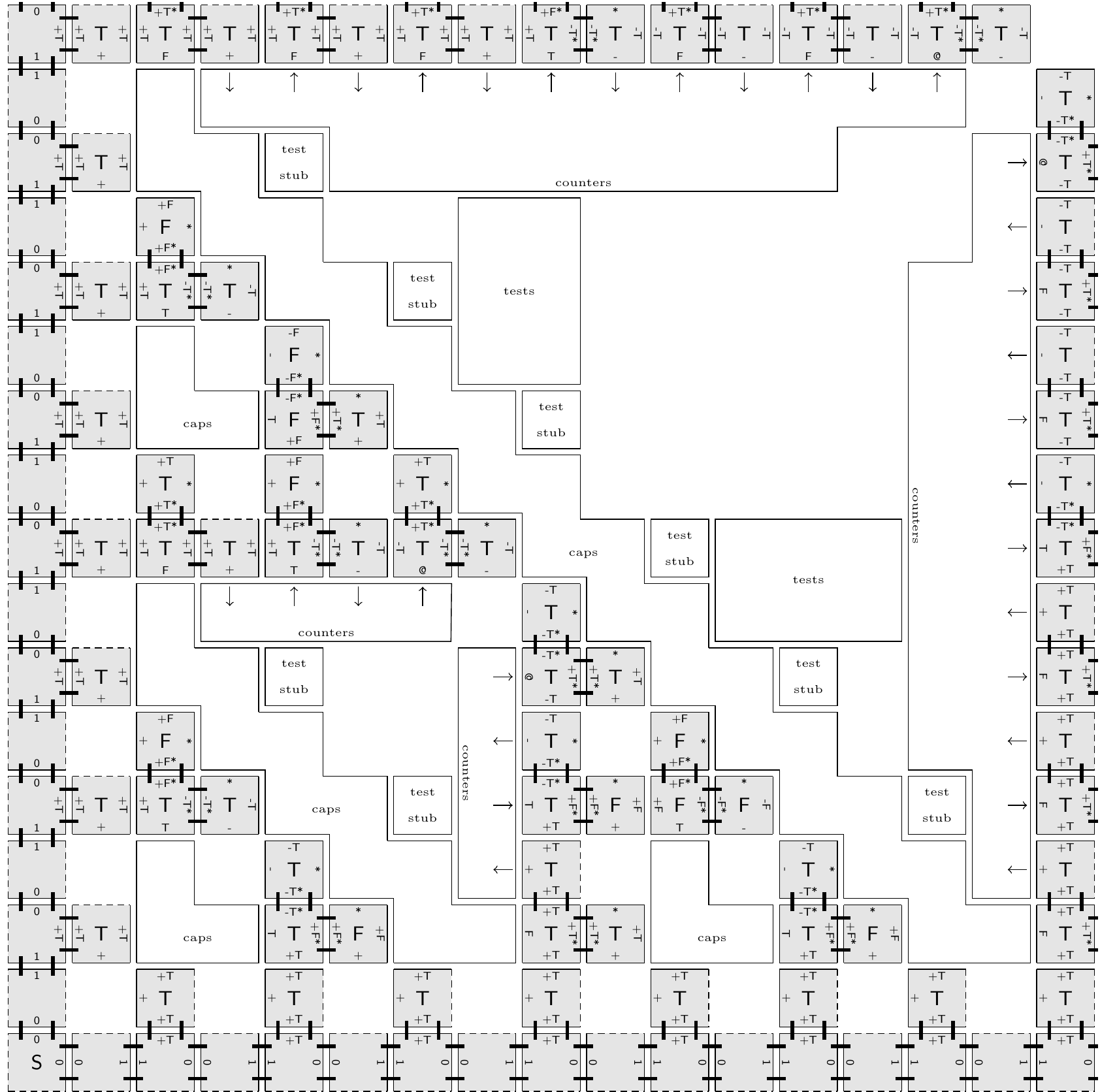}
	\caption{Example assembly of the horizontal and vertical bars of $\S$.}
	\label{fig:LacedExampleSBars}
\end{figure}

There are five tile types to assemble the boundary of $\S$, two for the bottom boundary and two for the left boundary.
The bottom boundary is assembled by a tile with a west glue color of $\scrim{0}$ attaching to the east side of $\scrib{S}$ and a tile with a west glue color of $\scrim{1}$ attaching to the east side of it.
This process continues ad infinitum. The left boundary assembles in a similar fashion.
See Figure \ref{fig:LacedExampleSBars} for an illustration.

\begin{figure}[!ht]
	\centering
    \begin{minipage}[b]{3.3in}
        \subfloat[Cap Fibers]{\includegraphics[width=3in]{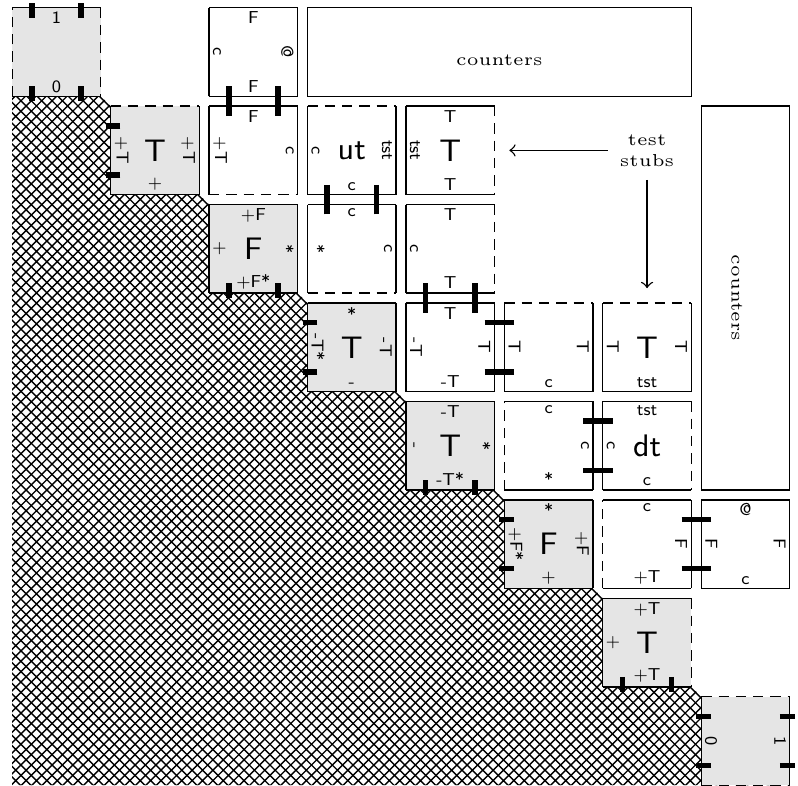}}\vspace{30pt}
        \subfloat[Test Fibers]{\includegraphics[width=3in]{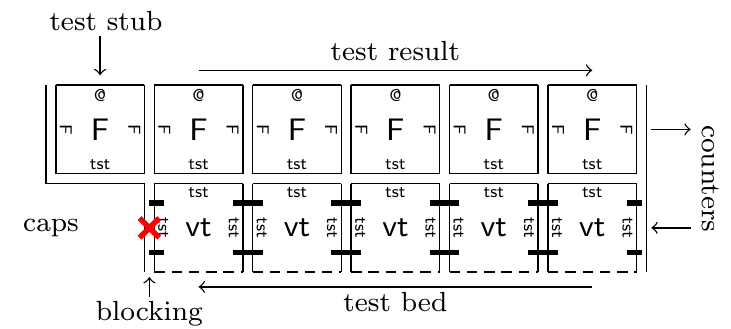}}\vspace{10pt}
    \end{minipage}\quad
    \subfloat[Vertical Bar and Counters]{\includegraphics[width=2in]{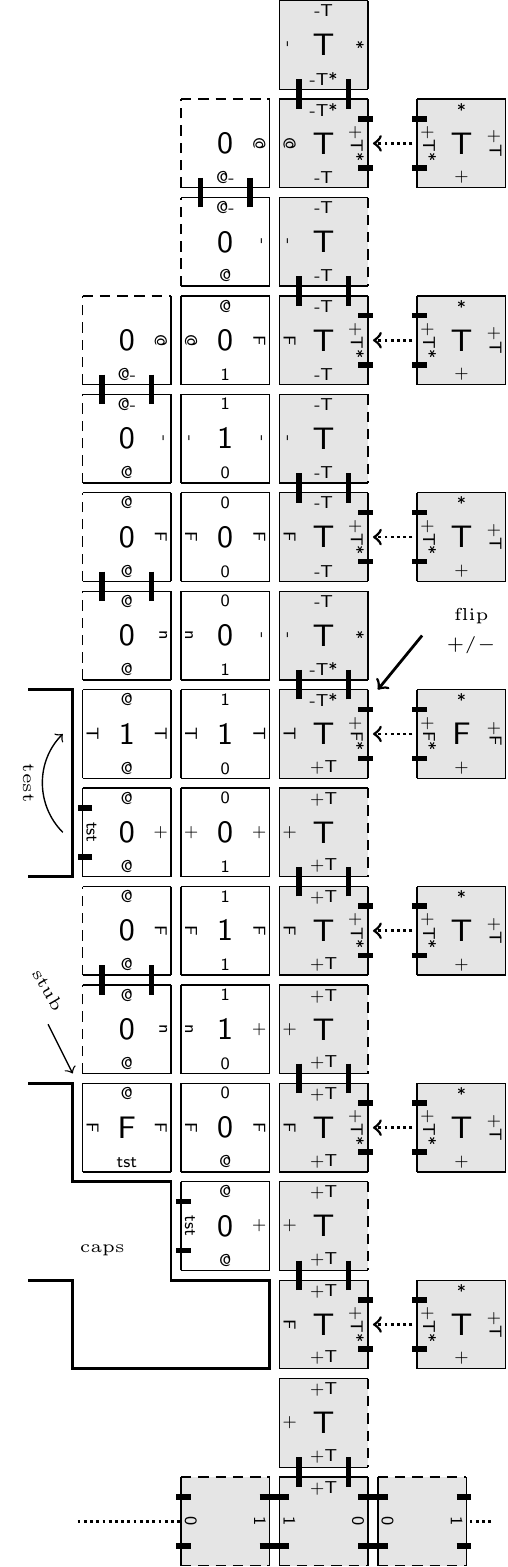}}
	\caption{Example assembly of the cap, test, and counter fibers in $\L$.}
	\label{fig:LacedExampleAssemblies}
\end{figure}

The are thirty-two tile types to assemble the horizontal and vertical bars in the interior of $\S$, sixteen for the vertical bars and sixteen for the horizontal bars.
Here, we focus on the assembly of a vertical bar.
A horizontal bar assembles in a similar fashion.
The glue colors on the north and south sides of the tiles in a vertical bar are made up of the characters $\scrim{+}, \scrim{-}, \scrim{*}, \scrim{T}$, and $\scrim{F}$.
Tiles used for the bottom half of a vertical bar use the $\scrim{+}$ and cause the counter that assembles next to the bar to increment.
Tiles used for the top half of a vertical bar use the $\scrim{-}$ and cause the counter to decrement.
A tile with a $\scrim{*}$ in its south glue color also has a $\scrim{*}$ for its east glue color.
The $\scrim{*}$ will be used by the cap tiles to know when to stop attaching.
The $\scrim{T}$ or $\scrim{F}$ in the north and south glue colors propagates through the entire bar.
When a location $\vm$ has a tile with a $\scrim{T}$ in the glue color of both its south and west sides it means that $\vm$ is the middle point of the largest diagonal to which it belongs and the cap tiles start their assembly at location $\vm$.
The west side of alternating tiles in the vertical bar have a glue color of either $\scrim{T}, \scrim{F}$, or $\scrim{@}$.
These glues allow the vertical bar to receive feedback from the counters.
A $\scrim{T}$ glue color tells the bar that it is at half of its intended height, at which point the bar switches from instructing the counter to increment to instructing the counter to decrement.
An $\scrim{@}$ glue color instructs the bar to complete its assembly.
An $\scrim{F}$ glue allows the bar continue assembling.
The east side of these tiles also starts the growth of new horizontal bars.
If the tile abuts a glue of color $\scrim{T}$ (or $\scrim{F}$) on the counter, then it negates this value for the horizontal bar originating on its east side. See Figures \ref{fig:LacedExampleSBars} and \ref{fig:LacedExampleAssemblies} for an illustration.

The cap tiles are fibers that sit on top of the diagonals of $\S$.
There are eighteen tile types to assemble the cap tiles.
Each diagonal of $\S$ has a vertical (horizontal) bar directly to the right (above) it.
The height (length) of these bars can be computed directly from knowledge of the location of the middle of this diagonal.
The test tiles will handle the two way communication between the caps and the counters so that this information can be used in the assembly of the bars.
The cap tiles also delay the assembly of these bars until the assembly of the relevant part of the diagonal has been completed.
This allows for the blocking behavior needed for proper assembly of these two-way communication fibers.
The middle point of the diagonal will always be at the unique location that has a tile with a $\scrim{T}$ in the glue color on both its south and west sides.
When tiles meeting this criteria are present, a cap tile will attach to the assembly at that location.
This will trigger the assembly of cap tiles both up and down the diagonal.
The growth of the cap tiles are controlled by the $\scrim{*}$ glue colors on the tops (right sides) of the vertical (horizontal) bars making up the diagonal.
They first attach to a tile having a $\scrim{*}$ in its glue color on the abutting side, and then to a tile having a $\scrim{T}$ or $\scrim{F}$ in its glue color on the abutting side, then the process repeats.
When a $\scrim{*}$ glue color is not present, the cap tiles cease their assembly.
See Figure \ref{fig:LacedExampleAssemblies} for an illustration.

There are thirty-four tile types to assemble the binary counter fibers that assemble adjacent to the bars of $\S$, seventeen for the vertical counters and seventeen for the horizontal counters.
The counters assemble in a zig-zag fashion.
Alternating rows go from east to west (zig) and west to east (zag).
The zig row either increments or decrements the value of the counter depending on the glue color on the abutting side of the tile on the bar of $\S$.
During the increment phase of the counter, if the counter is at a value that is one less than a power of two, then it initiates the two-way communication with the cap fibers by presenting a $\scrim{tst}$ glue.
The result of the test, either a $\scrim{T}$ or $\scrim{F}$ glue color, is propagated back to the bar.
If the counter is not at a power of two, then the zag row returns the value $\scrim{F}$.
During the decrement phase of the counter, if the counter is at a power of two, the zag row is triggered by a $\scrim{@-}$ glue color, which instructs the counter to shrink in width by one.
See Figure \ref{fig:LacedExampleAssemblies} for an illustration.

There are six tile types to assemble the test fibers. Three for the vertical tests and three for the horizontal tests.
These fibers allow for the two-way communication between the counter fibers and cap fibers.
The request made by the counter fibers is sent along the bottom row of the test fibers and the response is returned along the top row of the test fibers.
Because the counters do not assemble until the assembly of the corresponding caps have completed, we can be sure that the bottom row will not continue indefinitely -- it will be blocked by the cap fibers.
See Figure \ref{fig:LacedExampleAssemblies} for an illustration.
We conclude with the following theorems which prove that $\L$ strictly self-assembles in the Tile Assembly Model.

We now show that $\TASL$ satisifies the conditions for the generalization of local determinism we introduced in Section \ref{sec:cd}.

\begin{theorem}\label{thm:Lcd}
$\TASL$ is conditionally deterministic.
\begin{proof}
Let $\valpha=(\alpha_i \mid 0 \le i < k)$ be any assembly sequence in $\TASL$ such that $\res(\valpha)$ is terminal. It should be clear that there is such an assembly sequence and that $k = \infinity$.
First, we make the following observations that the reason that $\TASL$ is not locally deterministic is because of the locations in $\alpha$ at which there is a tile of type $\scrib{ut}$ or $\scrib{dt}$ (of Figure \ref{fig:LacedTiles}).
\begin{enumerate}
\item For each $0 \le i < k$, the unique tile type $t = \alpha(\vm)$, where $\vm \in \dom\alpha_{i+1}\setminus\dom\alpha_i$, attaches to $\alpha_i$ with a strength of exactly 2.
\item For each location  $\vm \in \dom\alpha \setminus \dom\alpha_0$ such that $\alpha(\vm)=\scrib{ut}$, either
\begin{equation*}
	\vm+\un \in \tfrontier{\scrib{ht}}\, (\alpha \upharpoonright (\dom\alpha \setminus (\set{\vm} \cup (\vm + \outsides(\vm)))))
\end{equation*}
or
\begin{equation*}
    \frontier (\alpha \upharpoonright (\dom\alpha \setminus (\set{\vm} \cup (\vm + \outsides(\vm))))) = \emptyset,
\end{equation*}
and for all $t \in \TL \setminus \set{\scrib{ut},\scrib{ht}}$,
\begin{equation*}
	\tfrontier{t}\, (\alpha \upharpoonright (\dom\alpha \setminus (\set{\vm} \cup (\vm + \outsides(\vm))))) = \emptyset.
\end{equation*}
\item For each location $\vm \in \dom\alpha \setminus \dom\alpha_0$ such that $\alpha(\vm)=\scrib{dt}$,
\begin{equation*}
	\vm+\ue \in \tfrontier{\scrib{vt}}\, (\alpha \upharpoonright (\dom\alpha \setminus (\set{\vm} \cup (\vm + \outsides(\vm)))))
\end{equation*}
or
\begin{equation*}
 \frontier (\alpha \upharpoonright (\dom\alpha \setminus (\set{\vm} \cup (\vm + \outsides(\vm))))) = \emptyset,
\end{equation*}
and all $t \in \TL \setminus \set{\scrib{ut},\scrib{ht}}$,
\begin{equation*}
	\tfrontier{t}\, (\alpha \upharpoonright (\dom\alpha \setminus (\set{\vm} \cup (\vm + \outsides(\vm))))) = \emptyset.
\end{equation*}
\item For each location $\vm \in \dom\alpha \setminus \dom\alpha_0$ such that $\alpha(\vm) \not\in \set{\scrib{ut},\scrib{dt}}$, and all $t \in \TL \setminus \set{\alpha(\vm)}$,
\begin{equation*}
	\tfrontier{t}\, (\alpha \upharpoonright (\dom\alpha \setminus (\set{\vm} \cup (\vm + \outsides(\vm))))) = \emptyset.
\end{equation*}
\item $\frontier \alpha = \emptyset$.
\end{enumerate}
Thus, $\valpha$ satisfies conditions (1) and (3) of both local determinism and conditional determinism. What prevents $\valpha$ from satisfying condition (2) of local determinism is the second and third observation above.
So, it suffices to show that
\begin{enumerate}
\item For each location $\vm \in \dom\alpha \setminus \dom\alpha_0$ such that $\alpha(\vm)=\scrib{ut}$,
\begin{equation*}
	\frontier (\alpha \upharpoonright (\dom\alpha \setminus (\set{\vm} \cup (\vm + (\outsides(\vm) \cup \depsides(\vm)))))) = \emptyset, \tand
\end{equation*}
\item For each location $\vm \in \dom\alpha \setminus \dom\alpha_0$ such that $\alpha(\vm)=\scrib{dt}$,
\begin{equation*}
	\frontier (\alpha \upharpoonright (\dom\alpha \setminus (\set{\vm} \cup (\vm + (\outsides(\vm) \cup \depsides(\vm)))))) = \emptyset.
\end{equation*}
\end{enumerate}
We will argue that (2) holds. The argument that (1) holds is similar. Let $\vm \in \dom\alpha \setminus \dom\alpha_0$ such that $\alpha(\vm)=\scrib{dt}$. By construction, it must be the case that either $\alpha(\vm + \ue)\uparrow$ or $\alpha(\vm + \ue)\downarrow$. If $\alpha(\vm + \ue)\uparrow$ then it follows that
\begin{align*}
    \frontier &(\dom\alpha \setminus (\set{\vm} \cup (\vm + \outsides(\vm))))) = \emptyset, \text{ so}\\
    \frontier &(\alpha \upharpoonright (\dom\alpha \setminus (\set{\vm} \cup (\vm + (\outsides(\vm) \cup \depsides(\vm)))))) = \emptyset.
\end{align*}
If $\alpha(\vm + \ue)\downarrow$ then the tile at $\alpha(\vm + \ue)$ must have attached along the bottom row of the test fibers initiated by the vertical bar directly to the right of $\vm$ $($i.e., $\alpha(\vm + \ue) = \scrib{vt})$. However, if you look at the assembly of this bar, the second tile from the bottom uses the bottom right location of these caps as an input side (see Figure \ref{fig:LacedExampleAssemblies} for an illustration). Thus, the vertical bar can not assemble above this point until all of the down caps along this diagonal have assembled. Thus, $\vm \prec_{\TASL} \vm+\ue$. Hence, $\vm+\ue \in \depsides(\vm)$. Thus,
\begin{align*}
\frontier (\alpha \upharpoonright (\dom\alpha \setminus (\set{\vm} \cup (\vm + (\outsides(\vm) \cup \depsides(\vm)))))) = \emptyset. & &\hfill \qedhere
\end{align*}
\end{proof}
\end{theorem}

We now show that $\L$ strictly self-assembles in $\TASL$.

\begin{theorem}\label{thm:Lstrict}
$\L$ strictly self-assembles in $\TASL$.
 \begin{proof}
We say some set of locations $X \subseteq \L$ \dquote{properly assembles} if the intended tile type was placed at each location in the set, and no tile is placed at a location in $\Ints[2] \setminus \L$.
By Theorem \ref{thm:Lcd} and Theorem \ref{thm:CDimpliesDirected}, $|\terminals[\TASL]|=1$. Pick the unique $\alpha\in\terminals[\TASL]$. It suffices to show that $\dom\alpha = \L$.
We make the following claims about $\TASL$.
\begin{itemize}
\item[(1)] If $\L[i]$ assembles properly, then $\S[i+1]$ assembles properly. To see this note that $\S[i] \subseteq \L[i]$. Then, the same mechanics used to assemble $\S[i]$ are used to assemble $(2^{i},0) + \S[i]$ and $(0,2^{i}) + \S[i]$. Then, by \eqref{eqn:Sstage}, $\S[i+1]$ assembles properly.
\item[(2)] If $\S[i]$ assembles properly, then $\Caps[i]$ assemble properly. To see this Let $\vm = (2^{i-1},2^{i-1})$. Note that the tallest (widest) vertical (horizontal) bar of $\S[i]$ originates from the boundary and hence propagates a $\scrim{T}$ glue color throughout its assembly. Then, $\vm + \us$ and $\vm + \uw$ have a glue color of $\scrim{T}$ on its $\un$ and $\ue$ sides respectively. Thus, $\Caps[i]$ are allowed to begin their assembly at location $\vm$ and since all of the smaller horizontal and vertical bars of $\S[i]$ assemble properly, the caps will assemble up and down the longest diagonal in $\S[i]$.
\item[(3)] If $\Caps[i]$ assembles properly, then the largest horizontal and vertical bar of $\S[i+1]$ along with $\Counters[i]$ and $\Tests[i]$ assemble properly. To see this note that the longest vertical and widest horizontal bar of $\S[i+1]$ cannot grow very far until $\Caps[i]$ has completed assembling (see Figure \ref{fig:LacedExampleAssemblies} for an illustration). At this point the proper assembly of these bars depends upon the proper assembly of $\Counters[i]$. But for $\Counters[i]$ to assemble properly only depends on $\Tests[i]$ to assemble properly, which in turn depends on $\Caps[i]$ to assemble properly.
\end{itemize}
Our proof by induction easily follows from these claims. It is easy to see that $\L[0], \L[1], \L[2], \L[3]$ properly assemble in $\TASL$. It suffices to show that if $\L[i]$ properly assembles in $\TASL$, then $\L[i+1]$ properly assembles in $\TASL$.
Suppose $\L[i]$ has properly assembled in $\TASL$. Then, by claim (1), $\S[i+1]$ assembles properly. Then, by claim (2), $\Caps[i+1]$ assemble properly. Then, by claim (3), $\Counters[i+1]$ and $\Tests[i+1]$ assemble properly. Hence $\L[i+1]$ assembles properly.
 \end{proof}
\end{theorem}

It is also interesting to note that $\S$ also weakly self-assembles in $\TASL$.

 \begin{observation}
 $\S$ weakly self-assembles in $\TASL$.
 \end{observation}
 
Instructions for simulating $\TASL$ with the ISU TAS \cite{isutas}
are available at \url{http://www.cs.iastate.edu/~shutters/asast} .

We conclude this section by presenting our second main theorem which shows that the bound given in our first main theorem, Theorem \ref{thm:close} is tight.
\begin{theorem}\label{thm:laced}
There exists a set $X\subseteq\Ints[2]$ with the following properties.
\begin{itemize}
\item[(1)] $\S \subseteq X$.
\item[(2)] $\ZetaDim(X\Delta\,\S)=\ZetaDim(\S)$.
\item[(3)] $X$ strictly self-assembles in the Tile Assembly Model.
\end{itemize}
\begin{proof}
Let $X=\L$. 
By Observation \ref{obs:SsubsetL}, condition (1) is satisfied. 
By Theorem \ref{thm:LZetaDim} and Observation \ref{obs:ZetaDimProps}, condition (2) is satisfied.
By Theorem \ref{thm:Lstrict} condition (3) is satisfied.
\end{proof}
\end{theorem}

%% file: open.tex

\section{Open Questions}\label{sec:open}

Our results show that in the case of the Sierpinski triangle, no set \dquote{close} to the Sierpinski triangle  strictly self-assembles.  Given that no self-similar fractal is known to strictly self-assemble, a natural question is whether there exists a self-similar fractal that can be approximated closely. Is there a set $X$ that strictly self-assembles and a self-similar fractal $F$ such that $\ZetaDim(X \Delta\, F)<\ZetaDim(F)$?

We\,demonstrated\,a\,distortion-free\,fibering\,technique\,that\,enables\,a\,superset of the Sierpinski\,triangle\,to\,strictly\,self-assemble\,without\,increasing\,the fractal dimension of the intended structure.
However, this technique depends on properties unique to the Sierpinski triangle and does not generalize to a large class of fractals. Is there a distortion-free fibering technique that generalizes to a large class of fractals without increasing the fractal dimension of the intended structure?

We gave an extension of local determinism sufficient for showing a blocking tile assembly system is directed.  However, the relative order of when certain tiles attach \emph{in every assembly sequence} must be established. In contrast, local determinism requires only the analysis of a single assembly sequence. Is there a set of conditions that only requires the analysis of a single assembly sequence that is sufficient for showing a blocking tile assembly system is directed?